\documentclass[%
 reprint,
 amsmath,amssymb,
 aps, pra,
 longbibliography,
 twocolumn
]{revtex4-1}

\usepackage{graphicx}
\usepackage{dcolumn}
\usepackage{bm}
\usepackage{siunitx}
\usepackage{nicefrac}
\usepackage{mathrsfs}
\usepackage{dsfont}
\usepackage{lipsum}
\usepackage{amsmath}
\usepackage{braket}  % for \ket and \bra commands
\usepackage[hidelinks]{hyperref}

\DeclareUnicodeCharacter{2009}{\,}

\begin{document}
	
\renewcommand{\vec}[1]{\mathbf{#1}}
\title{High-Fidelity Quantum State Transfer in Multimode Resonators via Tunable Pulses}

\author{Yuanning Chen}
\author{Xinxin Yang}
\author{Simon Gröblacher}
\email{s.groeblacher@tudelft.nl}

\affiliation{Kavli Institute of Nanoscience, Department of Quantum Nanoscience, Delft University of Technology, 2628CJ Delft, The Netherlands}

%\date{\today}

\begin{abstract}
Quantum state transfer between distant nodes is essential for distributed quantum information processing. Existing protocols are typically optimized for specific coupling regimes, such as adiabatic dark-state transfer in the single-mode limit and pitch-and-catch schemes in the multimode regime, leaving the crossover between them without a simple and unified control strategy. Here we identify a minimal two-parameter control framework that enables high-fidelity quantum state transfer across this single-mode-to-multimode crossover in a multimode quantum channel. Using a pulse-shaped pitch-and-catch protocol controlled only by the pulse ramp rate and the emission-absorption delay, we achieve transfer fidelities exceeding $99.9\%$, extending pitch-and-catch protocols toward the single-mode limit without requiring dark-state protection or complex pulse design. We further demonstrate robustness against dissipation, disorder, detuning, and imperfect initialization under experimentally realistic conditions. These results provide a simple and broadly applicable framework for state transfer in multimode quantum channels, with relevance to circuit-QED and hybrid quantum-acoustic systems.
\end{abstract}

\maketitle

\section{Introduction}

Practical quantum computing at scale will likely require millions of qubits distributed across many interconnected modules~\cite{nielsen2010quantum,fowler2012surface}. While current superconducting quantum processors integrate only hundreds of qubits on a single chip~\cite{gao2025establishing,google2025quantum}, modular architectures address this scalability challenge by coherently interconnecting smaller processing units using quantum state transfer (QST) and remote entanglement generation~\cite{axline2018demand, kurpiers2018deterministic, hermans2022qubit}. These capabilities are central to scalable superconducting quantum processors and the emerging quantum internet~\cite{wehner2018quantum, awschalom2021development}. However, achieving robust, high-fidelity QST over centimeter- to meter-scale distances remains challenging due to dissipation, decoherence, non-Markovian channel dynamics, and the absence of control protocols that operate reliably across different coupling regimes~\cite{breuer2016colloquium, he2025quantum}.

A representative superconducting circuit-QED architecture is shown schematically in Fig.~1(a): two qubits are coupled to a common resonator that serves as a quantum communication channel. The channel supports a set of discrete standing-wave modes separated by the free spectral range $\nu_{\mathrm{fsr}}$, and each qubit couples dynamically to these modes via a tunable coupler~\cite{chen2014qubit, yan2018tunable}. Depending on the ratio between the qubit--channel coupling strength $g$ and $\nu_{\mathrm{fsr}}$, qualitatively different state-transfer mechanisms emerge.

In the single-mode strong-coupling (SMSC) limit, $\nu_{\mathrm{fsr}} \gg g$ $\gg$ dissipation, only a single resonant mode participates in the dynamics. In this regime, QST can be implemented using adiabatic dark-state protocols analogous to stimulated Raman adiabatic passage (STIRAP)~\cite{vitanov2017stimulated, chang2020remote}. The Hamiltonian supports a dark eigenstate with negligible channel population, allowing coherent state transfer via adiabatic variation of the relative qubit--mode coupling strengths [Fig.~1(b)].

When the coupling strength becomes comparable to the free spectral range ($g \sim \nu_{\mathrm{fsr}}$), the qubits hybridize simultaneously with many resonator modes. This multimode strong-coupling (MMSC) regime can be realized using long coplanar resonators in circuit QED~\cite{sundaresan2015beyond, kuzmin2019superstrong} or by exploiting the slow group velocity of acoustic waves in hybrid quantum-acoustic systems~\cite{satzinger2018quantum, moores2018cavity, scigliuzzo2025quantum}. In this regime, state transfer is no longer well described by adiabatic eigenstate following, but instead relies on the controlled emission and reabsorption of itinerant wavepackets formed from coherent superpositions of multiple standing-wave modes. Time-symmetric emission and capture of shaped wavepackets—often termed pitch-and-catch—enable high-fidelity state transfer~\cite{cirac1997quantum, korotkov2011flying, sete2015robust, bienfait2019phonon, qiao2023splitting, grebel2024bidirectional}.

While the SMSC and MMSC limits are individually well understood, control in the intermediate crossover regime remains challenging. In this regime, dark eigenstates underlying adiabatic protocols are degraded by strong multimode hybridization, whereas wavepacket-based approaches suffer from dispersion and uncontrolled non-Markovian memory effects~\cite{vogell2017deterministic, he2025quantum}. Previous work has extended dark-state protocols to multimode systems~\cite{vogell2017deterministic,chang2020remote,malekakhlagh2024enhanced} and developed non-Markovian frameworks that bridge adiabatic and wavepacket regimes~\cite{he2025quantum}. However, these approaches typically require engineered eigenstate protection, auxiliary driving, or complex pulse design, increasing calibration complexity and making the transfer mechanism less transparent. As a result, a minimal and experimentally accessible protocol that operates efficiently across the multimode-to-single-mode crossover remains lacking.

In this work, we numerically study a pulse-shaped pitch-and-catch protocol that enables high-fidelity quantum state transfer in the crossover between the single-mode and multimode coupling regimes using two control parameters: the pulse ramp rate and the emission-absorption delay. By optimizing these parameters, we achieve transfer fidelities exceeding $99.9\%$, continuously extending pitch-and-catch protocols from the multimode regime toward the single-mode limit. Importantly, these two parameters admit direct physical interpretations: the ramp rate controls the degree of adiabaticity, while the delay captures the propagation time of the itinerant wavepacket. We further show that it remains robust under experimentally relevant conditions, including dissipation, disorder, detuning, imperfect initial states, and residual thermal excitations in the channel.

Rather than introducing additional control Hamiltonians or complex pulse engineering, our approach provides a simple control framework for designing state-transfer protocols in multimode channels. Because all relevant timescales scale naturally with the free spectral range $\nu_{\mathrm{fsr}}$, the results presented here offer a broadly applicable prescription for state transfer, with particular relevance to circuit-QED and hybrid quantum-acoustic architectures.

\begin{figure}[h!]
	\centering
	\includegraphics[width=0.48\textwidth]{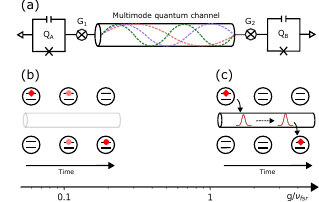}
	\caption{\textbf{State-transfer schematics.}
	(a) Two qubits (labelled as $Q_A$ and $Q_B$) are coupled to a common multimode resonator acting as a quantum channel. 
	Remote state transfer is implemented by dynamically tuning the qubit-channel coupling via tunable couplers (labelled $G_1$ and $G_2$). 
	(b) In the single-mode limit, the system supports a dark eigenstate with negligible population in the resonator. 
	A STIRAP-like adiabatic sequence transfers the state from $Q_A$ to $Q_B$ while avoiding excitation of the channel. 
	(c) In the multi-mode limit, the dark-state condition breaks down. 
	State transfer is achieved using the pitch-and-catch protocol: $Q_A$ emits a shaped photon/phonon wavepacket into the multimode resonator (pitch), and $Q_B$ time-reverses the process to reabsorb it (catch).}
	\label{fig:1}
\end{figure}

\section{Model}

The system consists of two qubits ($Q_A$ and $Q_B$) coupled to a multimode quantum channel via tunable couplers ($G_1$ and $G_2$), as illustrated in Fig.~1(a). The couplers allow time-dependent control of the interaction strengths $g_A(t)$ and $g_B(t)$. The quantum channel supports discrete standing-wave modes with frequencies $\omega_k$ separated by the free spectral range $\nu_{\mathrm{fsr}}$.

In the interaction picture and under the rotating-wave approximation (RWA), the effective Hamiltonian is~\cite{he2025quantum}
\begin{equation}
\begin{split}
H_I(t) = & \; g_A(t)\, \hat{\sigma}_A^\dagger \sum_{k} \hat{c}_k e^{-i \Delta_k^A t} \\
& + g_B(t)\, \hat{\sigma}_B^\dagger \sum_{k} (-1)^k \hat{c}_k e^{-i \Delta_k^B t} + \mathrm{H.c.}
\end{split}
\end{equation}

Here, $\hat{\sigma}_{A}$ and $\hat{\sigma}_{B}$ denote the lowering operators of $Q_A$ and $Q_B$, respectively, and $\hat{c}_{k}$ annihilates a photon (or phonon) in the $k$-th mode of the channel. The phase factor $(-1)^k$ accounts for the qubits’ approximate half-wavelength separation: adjacent standing-wave modes have alternating field parity, causing opposite coupling signs~\cite{he2025quantum,chang2020remote}. The detunings are defined as $\Delta_k^A = \omega_k - \omega_A$ and $\Delta_k^B = \omega_k - \omega_B$. The RWA is valid when the coupling strengths $g_{A,B}$ and detunings are small compared to the qubit and mode frequencies, i.e., $\omega_{A,B}, \omega_k \gg g_{A,B}, |\Delta_k^{A,B}|$.

Before addressing state transfer between two qubits, we first examine the dynamics when only $Q_A$ is coupled to the multimode channel ($g_B = 0$). We take $Q_A$ to be resonant with the central channel mode and include 51 resonator modes symmetrically around this frequency. This finite-mode truncation balances numerical efficiency and accuracy: the resulting bandwidth ($\sim 25\nu_{\mathrm{fsr}}$ on either side) greatly exceeds the largest coupling strength considered, ensuring that the emitted excitation remains spectrally confined and that edge-mode populations remain negligible throughout the dynamics~\cite{scigliuzzo2025quantum, calajo2016atom}. The system is initialized with a single excitation in $Q_A$.

In the absence of pulse shaping, i.e., for constant coupling $g$, the resulting dynamics are determined by the competition between the coupling strength $g$ and the free spectral range $\nu_{\mathrm{fsr}}$. In the single-mode limit, the excitation exchanges coherently between the qubit and a single cavity mode, giving rise to simple Rabi oscillations with period $1/2g$. In the multi-mode limit, the excitation spreads into several modes and returns to the qubit as pulsed revivals at time intervals $t \sim 1/\nu_{\mathrm{fsr}}$~\cite{milonni1983exponential,krimer2014route,meiser2006superstrong}. The ratio $g/\nu_{\mathrm{fsr}}$ therefore controls the crossover between localized Rabi exchange and itinerant wavepacket dynamics. An example in the crossover regime ($2g=\nu_{\mathrm{fsr}}$) is shown in Fig.~2(a), where the qubit population displays incomplete revivals with gradually decreasing amplitude. The corresponding frequency-resolved mode population in Fig.~2(b) reveals the origin of this behavior: emission into multiple modes causes dispersive broadening of the wavepacket, which is not effectively reversed during reabsorption due to the accumulation of multimode phase dispersion. As a result, each emission-catch cycle yields a more delocalized wavepacket that is weakly reabsorbed by the qubit.

Having established the characteristic dynamics under constant coupling, we then apply a pulse-shaped extension of the pitch-and-catch protocol to recover the qubit population. Specifically, we employ a hyperbolic-secant pulse with three segments,

\begin{equation}
g(t) = 
\begin{cases}
g_0 \,\mathrm{sech}\!\big[\kappa (t - \tau)\big], & 0 < t < \tau, \\[6pt]
g_0, & \tau \leq t < \tau + \tau_d, \\[6pt]
g_0 \,\mathrm{sech}\!\big[\kappa (t - (\tau + \tau_d))\big], & \tau + \tau_d \leq t < 2\tau + \tau_d,
\end{cases}
\end{equation}

where $\tau = 6/\kappa$ defines the ramp duration, $\kappa$ is the ramp rate, and $\tau_d$ is the inter-pulse delay, i.e., the holding time at maximum coupling $g_0$. 

We emphasize that the specific choice of the hyperbolic-secant pulse is not essential to the protocol. Rather, its use is motivated by two generic features: (i) the smooth turn-on and turn-off suppress non-adiabatic spectral leakage during emission, and (ii) the approximate time-reversal symmetry of the pulse facilitates efficient reabsorption of the returning wavepacket~\cite{bienfait2019phonon, qiao2023splitting}. Similar performance can be achieved with other smooth, band-limited pulse shapes, such as Gaussian or raised-cosine ramps. The present choice therefore serves as a concrete and experimentally realistic pulse family that allows the protocol to be characterized using only two control parameters. 

When $\kappa$ and $\tau_d$ are chosen appropriately, Figs.~2(c-d) show that the qubit excitation is fully recovered after one round trip. The frequency-resolved mode dynamics yield two observations. First, the shaped pulse significantly suppresses spectral broadening during emission, in contrast to the constant-coupling case shown in Fig.~2(b). Second, the symmetric pulse enforces a reversed spectral evolution during capture, effectively unwinding the modal phases acquired during emission and restoring the wavepacket to the qubit without distortion.

\begin{figure}[h!]
	\centering
	\includegraphics[width=0.48\textwidth]{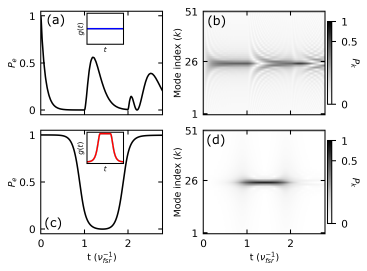}
	\caption{\textbf{Pulse shaping suppresses multimode dispersion and enables high capture efficiency.}
	We consider an intermediate coupling ratio ($g/\nu_{\mathrm{fsr}} = 0.5$) and examine single-qubit round-trip dynamics by setting $g_B = 0$, so that only $Q_A$ interacts with the multimode channel.  
	Top (bottom) panels show constant coupling (sech-shaped pulses), including qubit population and mode-resolved dynamics.
	(a)~With constant coupling, the excitation emitted by $Q_A$ is only weakly reabsorbed, resulting in incomplete and progressively weaker revivals over successive round trips.  
	(b)~The corresponding modal population shows strong multimode dispersion and clear asymmetry between emission and capture, leading to wavepacket delocalization and poor reabsorption.  
	(c)~With shaped coupling $g_A(t)$, $Q_A$ fully reabsorbs the returning wavepacket, yielding near-unit return probability.  
	(d)~The shaped pulse suppresses multimode dispersion and enforces time-reversal symmetry between emission and capture, enabling complete reabsorption of the wavepacket.}
	\label{fig:2}
\end{figure}

Having characterized the single-qubit dynamics, we now activate the second coupling
$g_B(t)$ to perform quantum state transfer from $Q_A$ to $Q_B$. Both qubits are tuned
into mutual resonance and resonant with the central resonator mode. The system is
initialized with a single excitation in $Q_A$, and the objective is to transfer this
excitation to $Q_B$ by extending the pitch-and-catch protocol from the multi-mode limit
into the crossover toward the single-mode limit.

For numerical simplicity, we assume equal maximum coupling strengths ($g_A^{\rm max} = g_B^{\rm max} = g$) and employ identical hyperbolic-secant ramps for $g_A(t)$ and $g_B(t)$~\cite{zhong2019violating}. The protocol is therefore governed by only two independent control parameters: (i) the normalized ramp rate $\kappa / \nu_{\mathrm{fsr}}$, and (ii) the normalized delay $\tau_d \nu_{\mathrm{fsr}}$ between the emission and absorption ramps.

For each coupling ratio $g/\nu_{\mathrm{fsr}}$, we sweep the two-dimensional parameter space $(\kappa, \tau_d)$ to maximize the transfer fidelity $F$, defined as the final population of qubit~B, $P_B(t_{\mathrm{cycle}})$. For clarity, we plot the infidelity $1-F$ throughout. Representative colormaps for $g/\nu_{\mathrm{fsr}} = 0.2$ and $0.8$ are shown in Fig.~3(a,b), where red markers indicate the optimal parameters. We note that Fig.~3(b) also shows a high-fidelity region at small $\kappa$. This branch corresponds to a longer protocol duration, where the excitation undergoes multiple back-and-forth transfers rather than a single emission-absorption event. Here we focus on the fastest high-fidelity single-transfer process. The extracted optimal values
are summarized in Fig.~3(c-d), together with empirical fits, revealing an approximately
exponential dependence of the optimal ramp rate $\kappa$ (and the total cycle time
$\tau_{\mathrm{cycle}}$) and a logarithmic dependence of the optimal delay $\tau_d$ on
$g/\nu_{\mathrm{fsr}}$. Figure~3(e) shows the corresponding time-domain dynamics under
optimal conditions for representative coupling ratios. The solid curves denote the transfer
infidelity $1 - P_B(t)$, while the dashed curves show the residual infidelity on qubit~A,
$1 - P_A(t)$, both evaluated using the optimal parameters extracted from
Fig.~3(c-d).

\begin{figure}[h!]
	\centering
	\includegraphics[width=0.48\textwidth]{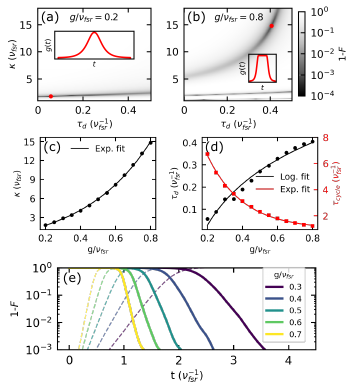}
	\caption{\textbf{State transfer at arbitrary interaction strength.}
	We optimize the pitch-and-catch protocol to transfer a single excitation from $Q_A$ to $Q_B$ in the crossover regime ($g/\nu_{\mathrm{fsr}}\in[0.2,0.8]$) by tuning two control parameters: the ramp speed $\kappa$ and the delay $\tau_d$ between the emission and absorption ramps.
	(a,b)~Infidelity $1-F$ as a function of $(\kappa,\tau_d)$ for $g/\nu_{\mathrm{fsr}}=0.2$ and $g/\nu_{\mathrm{fsr}}=0.8$, respectively. Red markers indicate the optimal points that maximize the transfer fidelity $F$, with the insets showing the corresponding ramp functions.
	(c,d)~Optimal parameters extracted from two-dimensional scans across multiple coupling ratios. The optimal ramp speed $\kappa$, optimal delay $\tau_d$, and total cycle duration $\tau_{\rm cycle}$ are plotted as functions of $g/\nu_{\mathrm{fsr}}$, together with exponential and logarithmic fits (solid lines).
	(e)~Time-domain dynamics under optimal conditions for representative values of $g/\nu_{\mathrm{fsr}}$. Solid curves show the transfer infidelity $1-P_B(t)$, while dashed curves show $1-P_A(t)$, the inverse of the residual population on $Q_A$.}
	\label{fig:3}
\end{figure}

Remarkably, the pulse-shaped pitch-and-catch protocol achieves near-unity transfer fidelity ($F > 99.9\%$) across the crossover regime using a single pulse family with only two tunable parameters. The optimal parameter trends reveal two distinct limiting behaviors. In the vicinity of the single-mode limit [Fig.~3(a)], the transfer is primarily sensitive to the ramp rate $\kappa$ and only weakly dependent on the delay $\tau_d$. Optimal performance requires slow ramps (small $\kappa$), such that the system evolves quasi-adiabatically, while $\tau_d$ becomes short or negligible. In this regime, the dynamics deviate from the conventional pitch-and-catch picture and instead resemble a STIRAP-like adiabatic process. In contrast, close to the multi-mode limit [Fig.~3(b)], the dynamics are dominated by itinerant wavepacket transport. Here, the transfer becomes more sensitive to the delay $\tau_d$, which sets the propagation time of the wavepacket, while the dependence on the ramp rate $\kappa$ is comparatively weaker. Optimal performance is obtained with fast ramps (large $\kappa$) and longer delays, with the total transfer duration approaching $1/(2\nu_{\mathrm{fsr}})$. This scaling reflects the propagation of the emitted wavepacket: while $1/\nu_{\mathrm{fsr}}$ corresponds to a round trip in the channel, the state-transfer protocol requires only a single traversal. 

Although we do not derive analytical expressions for the optimal ramp rate $\kappa$ and delay $\tau_d$, their dependence on $g/\nu_{\mathrm{fsr}}$, as shown in Fig.~3(c,d), captures the crossover from localized, adiabatic dynamics in the single-mode limit to wavepacket-mediated transport in the multi-mode limit. The empirical fits further provide a practical guideline for parameter selection: once calibrated at a few representative coupling strengths, optimal values at nearby regimes can be obtained by interpolation, reducing the need for exhaustive parameter scans. However, this interpolation breaks down deep in the single-mode limit, where $\tau_d$ cannot be reduced below zero.

Two points regarding the scope of the protocol are worth noting. First, the robustness of
the protocol originates from approximate time-reversal symmetry between the emission and
absorption processes. As a result, both amplitude and phase information are preserved during
transfer, and the protocol remains effective for arbitrary superpositions within the qubit
$\{|g\rangle,|e\rangle\}$ manifold. Second, the optimization presented here targets the
fundamental $|g\rangle\!\leftrightarrow\!|e\rangle$ transition. For multilevel systems such
as transmons, state transfer involving higher excited states constitutes a distinct
scattering problem due to anharmonicity-induced detunings (e.g., between
$|e\rangle\!\leftrightarrow\!|f\rangle$ and $|g\rangle\!\leftrightarrow\!|e\rangle$
transitions) and different coupling matrix elements, and would therefore require
re-optimization of the pulse parameters.

Overall, the pulse shaping considered here provides a minimal and experimentally realistic
control strategy. While higher fidelities may be achievable using more elaborate pulse
shapes or additional control parameters, the simplicity of the present approach—based on a
single smooth pulse family and only two tunable parameters—makes it directly applicable to
existing circuit-QED and hybrid quantum-acoustic platforms with tunable couplers
\cite{chen2014qubit, yan2018tunable}, and enables extension of pitch-and-catch protocols
beyond the purely multimode regime.

\section{Robustness Under System Imperfections}

Having established the optimal coherent protocol, we now test its tolerance to realistic device imperfections. 
We assess the transfer fidelity in the presence of qubit relaxation, channel loss, multimode frequency disorder, and qubit-qubit detuning. 
In all cases, the control pulses are fixed to the optimal ramp parameters extracted in Fig.~3(c-d).

Dissipation is incorporated through qubit relaxation at rate $\gamma$ and resonator decay at rate $\kappa_c$, using the Lindblad master equation
\begin{equation}
\frac{d\rho}{dt} = -\,i[H_I(t),\rho]
+ \gamma\sum_{j\in\{A,B\}}\mathcal{D}[\hat{\sigma}_j]\rho
+ \kappa_c\sum_{k}\mathcal{D}[\hat{c}_k]\rho,
\end{equation}
where $\mathcal{D}[\hat{L}]\rho = \hat{L}\rho\hat{L}^\dagger - \tfrac12\{\hat{L}^\dagger\hat{L},\rho\}$. 
Figures~4(a) and 4(b) show the infidelity as a function of $\gamma/\nu_{\mathrm{fsr}}$ and $\kappa_c/\nu_{\mathrm{fsr}}$, respectively. 
We sweep $\gamma/\nu_{\mathrm{fsr}}$ from $10^{-5}$ to $10^{-2}$ and $\kappa_c/\nu_{\mathrm{fsr}}$ from $10^{-5}$ to $10^{-1}$, covering the parameter ranges relevant to current superconducting-circuit and quantum-acoustic systems. 
For example, transmon qubits typically exhibit $T_1$ times from a few microseconds up to several hundred microseconds, while microwave and phononic resonators can exhibit linewidths ranging from tens of kilohertz to a few megahertz, with $\nu_{\mathrm{fsr}}$ spanning from a few to hundreds of megahertz depending on geometry~\cite{kjaergaard2020superconducting,blais2021circuit,moores2018cavity,bienfait2019phonon}.
Across these realistic ranges the protocol maintains high fidelity, with infidelity generally remaining below $10^{-2}$. 
Larger values of $g/\nu_{\mathrm{fsr}}$ are naturally more tolerant due to shorter transfer durations.

We next consider fabrication-induced frequency disorder in the multimode channel, modeled as static Gaussian fluctuations of the resonator mode frequencies with standard deviation $\sigma=\delta\,\nu_{\mathrm{fsr}}/3$, such that nearly all modes lie within $\pm\delta\,\nu_{\mathrm{fsr}}$. Small deviations from uniform mode spacing can arise from geometric inhomogeneity, coupling-induced boundary conditions, and lithographic disorder, and have been observed in multimode spectra of both superconducting and acoustic platforms~\cite{sundaresan2015beyond,zhang2017multimode,zivari2022chip}. Figure~4(c) shows the disorder-averaged infidelity (100 realizations) as a function of $\delta$. The transfer fidelity remains high across the experimentally relevant range. The multi-mode limit is more tolerant of disorder because the transfer involves the collective hybridization of many modes and therefore partially averages over random frequency offsets, whereas in the single-mode limit the dynamics rely on a single resonant mode and are correspondingly more sensitive to frequency perturbations.

Finally, we examine qubit-qubit frequency mismatch by introducing a static detuning $\Delta\omega$ between the two qubits. 
Small offsets of order $1$–$5~\mathrm{MHz}$, typical in tunable transmon devices due to flux noise and calibration drift~\cite{kjaergaard2020superconducting}, have negligible influence on the fidelity, as shown in Fig.~4(d). 
The protocol remains robust for $|\Delta\omega| \lesssim 10^{-2}\nu_{\mathrm{fsr}}$, where the infidelity stays below $10^{-3}$. 
Larger detunings eventually degrade the transfer, but the experimentally relevant range lies well within the tolerant regime. 
As before, the multi-mode limit accommodates larger detunings due to its broader hybridization bandwidth, while the single-mode limit is more sensitive because it relies on a sharp qubit-mode resonance.

\begin{figure}[h!]
	\centering
	\includegraphics[width=0.48\textwidth]{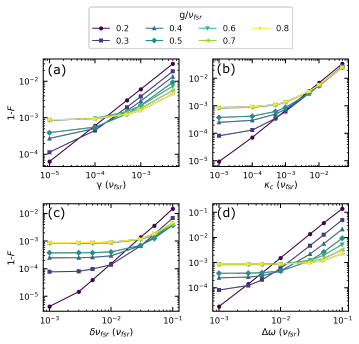}
	\caption{\textbf{Robustness of state transfer against dissipation, disorder, and frequency mismatch.}
	Transfer infidelity $1-F$ for different coupling ratios $g/\nu_{\mathrm{fsr}}$ in the presence of realistic imperfections.
	(a) Infidelity versus qubit decay rate $\gamma$.
	(b) Infidelity versus resonator decay rate $\kappa_c$.
	(c) Infidelity versus mode-frequency disorder, with Gaussian fluctuations of standard deviation $\sigma=\delta\nu_{\mathrm{fsr}}/3$ (data averaged over 100 realizations).
	(d) Infidelity versus qubit frequency mismatch $\Delta\omega$.
	In the low-loss, low-disorder, and small-detuning limits, the curves saturate at the intrinsic unitary infidelity obtained from Fig.~3, which exceeds 99.9\% and is higher in the single-mode limit. 
	Across all imperfections, the protocol remains most robust in the multi-mode limit due to the shorter transfer time and larger hybridization bandwidth.}
	\label{fig:4}
\end{figure}

\section{Robustness to Imperfect Initial States}
\label{sec:imperfections}

In the previous section we assumed an ideal initial state, where the qubit is perfectly prepared in $\ket{e}$, the channel is in the vacuum state, and the dynamics are restricted to the single-excitation manifold of two-level systems. In practice, however, superconducting artificial atoms are multilevel systems, and state preparation as well as the channel initialization are subject to finite errors. In this section we examine how such nonideal initial conditions affect the transfer fidelity. The control pulses are kept fixed at the values optimized for the ideal two-level, vacuum-channel case, and only the initial state is varied.

Throughout this analysis we remain within the rotating-wave approximation, so that the total excitation number is conserved. Imperfections are therefore modeled as classical mixtures in the initial density matrix rather than as coherent transitions between excitation manifolds.

To capture leakage to higher levels during state preparation, we model $Q_A$ as a three-level system and initialize the system in the mixed state
\begin{equation} 
\rho(0) = [(1-\epsilon_f)\ket{e}\!\bra{e} + \epsilon_f\ket{f}\!\bra{f}] \otimes \ket{\mathrm{vac}}\!\bra{\mathrm{vac}}.
\end{equation}

Such leakage can arise from finite pulse selectivity, residual thermal population, or measurement-based reset errors. In state-of-the-art superconducting platforms, leakage outside the computational subspace is typically at or below the $10^{-3}$ level for calibrated single-qubit operations~\cite{chen2016measuring,rol2019fast}.

The subsequent evolution is computed using the pulse parameters that maximize the ideal two-level transfer fidelity. Because the dynamics are linear in the density matrix, the final excited-state population of qubit~B is given by a weighted sum of contributions from the $\ket{e}$ and $\ket{f}$ manifolds. The resulting infidelity $1-P_{B,e}$ as a function of the preparation error probability $\epsilon_f$ is shown in Fig.~5(a) for $\alpha/\nu_{\mathrm{fsr}}=1$ and $10$, where $\alpha$ is the qubit anharmonicity spanning experimentally relevant regimes. For small $\epsilon_f$ the curves are dominated by the intrinsic two-level infidelity, while for larger $\epsilon_f$ the infidelity increases approximately linearly, reflecting the first-order sensitivity to population outside the computational subspace. For experimentally realistic leakage levels $\epsilon_f \lesssim 10^{-3}$, the transfer fidelity remains essentially unchanged.

We next consider the effect of a stray thermal or spurious photon in the multimode channel. For a thermal occupation $n_{\mathrm{th}}\ll 1$, the probability of having more than one photon is $\mathcal{O}(n_{\mathrm{th}}^2)$, so the dominant correction comes from single-photon contamination~\cite{clerk2010introduction}. We therefore initialize the system in the mixed state
\begin{equation}
\rho(0) =
\ket{e}\!\bra{e}_A \otimes \ket{g}\!\bra{g}_B \otimes \rho_{\mathrm{ch}},
\end{equation}
with
\begin{equation}
\rho_{\mathrm{ch}} =
(1-\epsilon_p)\ket{\mathrm{vac}}\!\bra{\mathrm{vac}}
+ \epsilon_p \sum_{k} w_k \ket{1_k}\!\bra{1_k},
\end{equation}
where $\ket{1_k}$ denotes a single photon in channel mode $k$, $\sum_k w_k = 1$, and for simplicity we take a uniform distribution $w_k = 1/N$ over the $N$ modes. The final population $P_{B,e}$ is obtained by averaging over all such single-photon configurations.

The resulting infidelity is shown in Fig.~5(b). As in the case of
preparation errors, the leading correction scales linearly with the
stray-photon probability $\epsilon_p$. For a 5$-$7~GHz resonator at
10$-$20~mK the thermal occupation lies in the range
$\bar{n}\sim10^{-5}$--$10^{-3}$, with state-of-the-art experimental
upper bounds as low as $\bar{n}\lesssim2\times10^{-4}$
\cite{wang2019cavity}. In this regime the transfer fidelity remains
limited by the intrinsic two-level performance.

These results show that pulse parameters optimized for the ideal two-level, vacuum-channel model remain effective under realistic state-preparation and thermal conditions, and that no additional reoptimization is required for experimentally relevant error levels.

\begin{figure}[t]
	\centering
	\includegraphics[width=0.48\textwidth]{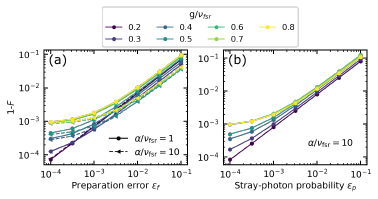}
	\caption{\textbf{Robustness to imperfect initial states.}
	(a) Transfer infidelity as a function of the preparation error probability $\epsilon_f$, where qubit~A is initialized in the mixed state $(1-\epsilon_f)\ket{e}\bra{e}+\epsilon_f\ket{f}\bra{f}$. Solid and dashed lines correspond to $\alpha/\nu_{\mathrm{fsr}}=1$ and $10$, respectively.
	(b) Transfer infidelity in the presence of a residual single photon in the channel with probability $\epsilon_p$, averaged uniformly over all channel modes. In both panels the control pulses are optimized for the ideal two-level, vacuum-channel case. The linear increase at large error probability reflects the first-order sensitivity to nonideal initial conditions, while the saturation at small error probability is set by the intrinsic two-level infidelity.}
	\label{fig:5}
\end{figure}
\section{Conclusion}

We have numerically demonstrated that a two-parameter pitch-and-catch protocol enables high-fidelity quantum state transfer across the crossover between single-mode and multi-mode limits in a multimode quantum channel. By optimizing the ramp rate and the emission-absorption delay, the protocol continuously interpolates between STIRAP-like dynamics in the single-mode limit and wavepacket-mediated transfer in the multi-mode limit, achieving transfer fidelities exceeding $99.9\%$.

The protocol remains robust against dissipation, disorder, detuning, and imperfect initial states. Its simplicity—requiring no auxiliary driving or complex pulse engineering—makes it well suited for scalable modular quantum architectures based on tunable couplers, including superconducting circuit-QED and hybrid quantum-acoustic platforms.

Looking ahead, several directions remain open. First, further analytical understanding of
the single-mode-to-multi-mode crossover would complement the present numerical study,
including analytical estimates for optimal pulse parameters and their scaling for general
pulse shapes. Second, our analysis is restricted to the excitation-conserving regime;
extending quantum state transfer into the ultrastrong-coupling domain, where the rotating-wave
approximation breaks down, remains an open challenge~\cite{frisk2019ultrastrong, egger2013multimode}.
Experimentally, the pulse-shaping protocol presented here is directly accessible with
existing hardware, and future work may explore refined pulse families based on optimal
control techniques~\cite{khaneja2005optimal, machnes2011comparing}, data-driven optimization
strategies~\cite{sivak2022model, fosel2018reinforcement, bukov2018reinforcement}, and
integration into larger-scale modular quantum-network architectures
\cite{wehner2018quantum, awschalom2021development,niu2023low}.

\section*{Acknowledgments}
We would like to thank Johannes Borregaard and Marius Villiers for insightful discussions and comments on the manuscript. This work is financially supported by the European Research Council (ERC CoG Q-ECHOS, 101001005) and is part of the research program of the Netherlands Organization for Scientific Research (NWO), supported by the NWO Frontiers of Nanoscience program, as well as through the Quantum Limits (SUMMIT.1.016) grant. This research is supported by The Kavli Foundation.


\begin{thebibliography}{49}%
	\makeatletter
	\providecommand \@ifxundefined [1]{%
		\@ifx{#1\undefined}
	}%
	\providecommand \@ifnum [1]{%
		\ifnum #1\expandafter \@firstoftwo
		\else \expandafter \@secondoftwo
		\fi
	}%
	\providecommand \@ifx [1]{%
		\ifx #1\expandafter \@firstoftwo
		\else \expandafter \@secondoftwo
		\fi
	}%
	\providecommand \natexlab [1]{#1}%
	\providecommand \enquote  [1]{``#1''}%
	\providecommand \bibnamefont  [1]{#1}%
	\providecommand \bibfnamefont [1]{#1}%
	\providecommand \citenamefont [1]{#1}%
	\providecommand \href@noop [0]{\@secondoftwo}%
	\providecommand \href [0]{\begingroup \@sanitize@url \@href}%
	\providecommand \@href[1]{\@@startlink{#1}\@@href}%
	\providecommand \@@href[1]{\endgroup#1\@@endlink}%
	\providecommand \@sanitize@url [0]{\catcode `\\12\catcode `\$12\catcode
		`\&12\catcode `\#12\catcode `\^12\catcode `\_12\catcode `\%12\relax}%
	\providecommand \@@startlink[1]{}%
	\providecommand \@@endlink[0]{}%
	\providecommand \url  [0]{\begingroup\@sanitize@url \@url }%
	\providecommand \@url [1]{\endgroup\@href {#1}{\urlprefix }}%
	\providecommand \urlprefix  [0]{URL }%
	\providecommand \Eprint [0]{\href }%
	\providecommand \doibase [0]{http://dx.doi.org/}%
	\providecommand \selectlanguage [0]{\@gobble}%
	\providecommand \bibinfo  [0]{\@secondoftwo}%
	\providecommand \bibfield  [0]{\@secondoftwo}%
	\providecommand \translation [1]{[#1]}%
	\providecommand \BibitemOpen [0]{}%
	\providecommand \bibitemStop [0]{}%
	\providecommand \bibitemNoStop [0]{.\EOS\space}%
	\providecommand \EOS [0]{\spacefactor3000\relax}%
	\providecommand \BibitemShut  [1]{\csname bibitem#1\endcsname}%
	\let\auto@bib@innerbib\@empty
	%</preamble>
	\bibitem [{\citenamefont {Nielsen}\ and\ \citenamefont
		{Chuang}(2010)}]{nielsen2010quantum}%
	\BibitemOpen
	\bibfield  {author} {\bibinfo {author} {\bibfnamefont {Michael~A}\
			\bibnamefont {Nielsen}}\ and\ \bibinfo {author} {\bibfnamefont {Isaac~L}\
			\bibnamefont {Chuang}},\ }\href@noop {} {\emph {\bibinfo {title} {Quantum
				computation and quantum information}}}\ (\bibinfo  {publisher} {Cambridge
		university press},\ \bibinfo {year} {2010})\BibitemShut {NoStop}%
	\bibitem [{\citenamefont {Fowler}\ \emph {et~al.}(2012)\citenamefont {Fowler},
		\citenamefont {Mariantoni}, \citenamefont {Martinis},\ and\ \citenamefont
		{Cleland}}]{fowler2012surface}%
	\BibitemOpen
	\bibfield  {author} {\bibinfo {author} {\bibfnamefont {Austin~G}\
			\bibnamefont {Fowler}}, \bibinfo {author} {\bibfnamefont {Matteo}\
			\bibnamefont {Mariantoni}}, \bibinfo {author} {\bibfnamefont {John~M}\
			\bibnamefont {Martinis}}, \ and\ \bibinfo {author} {\bibfnamefont {Andrew~N}\
			\bibnamefont {Cleland}},\ }\bibfield  {title} {\enquote {\bibinfo {title}
			{Surface codes: Towards practical large-scale quantum computation},}\
	}\href@noop {} {\bibfield  {journal} {\bibinfo  {journal} {Physical Review
				A—Atomic, Molecular, and Optical Physics}\ }\textbf {\bibinfo {volume}
			{86}},\ \bibinfo {pages} {032324} (\bibinfo {year} {2012})}\BibitemShut
	{NoStop}%
	\bibitem [{\citenamefont {Gao}\ \emph {et~al.}(2025)\citenamefont {Gao},
		\citenamefont {Fan}, \citenamefont {Zha}, \citenamefont {Bei}, \citenamefont
		{Cai}, \citenamefont {Cai}, \citenamefont {Cao}, \citenamefont {Chen},
		\citenamefont {Chen}, \citenamefont {Chen} \emph
		{et~al.}}]{gao2025establishing}%
	\BibitemOpen
	\bibfield  {author} {\bibinfo {author} {\bibfnamefont {Dongxin}\ \bibnamefont
			{Gao}}, \bibinfo {author} {\bibfnamefont {Daojin}\ \bibnamefont {Fan}},
		\bibinfo {author} {\bibfnamefont {Chen}\ \bibnamefont {Zha}}, \bibinfo
		{author} {\bibfnamefont {Jiahao}\ \bibnamefont {Bei}}, \bibinfo {author}
		{\bibfnamefont {Guoqing}\ \bibnamefont {Cai}}, \bibinfo {author}
		{\bibfnamefont {Jianbin}\ \bibnamefont {Cai}}, \bibinfo {author}
		{\bibfnamefont {Sirui}\ \bibnamefont {Cao}}, \bibinfo {author} {\bibfnamefont
			{Fusheng}\ \bibnamefont {Chen}}, \bibinfo {author} {\bibfnamefont {Jiang}\
			\bibnamefont {Chen}}, \bibinfo {author} {\bibfnamefont {Kefu}\ \bibnamefont
			{Chen}},  \emph {et~al.},\ }\bibfield  {title} {\enquote {\bibinfo {title}
			{Establishing a new benchmark in quantum computational advantage with
				105-qubit zuchongzhi 3.0 processor},}\ }\href@noop {} {\bibfield  {journal}
		{\bibinfo  {journal} {Physical Review Letters}\ }\textbf {\bibinfo {volume}
			{134}},\ \bibinfo {pages} {090601} (\bibinfo {year} {2025})}\BibitemShut
	{NoStop}%
	\bibitem [{goo(2025)}]{google2025quantum}%
	\BibitemOpen
	\bibfield  {title} {\enquote {\bibinfo {title} {Quantum error correction
				below the surface code threshold},}\ }\href@noop {} {\bibfield  {journal}
		{\bibinfo  {journal} {Nature}\ }\textbf {\bibinfo {volume} {638}},\ \bibinfo
		{pages} {920--926} (\bibinfo {year} {2025})}\BibitemShut {NoStop}%
	\bibitem [{\citenamefont {Axline}\ \emph {et~al.}(2018)\citenamefont {Axline},
		\citenamefont {Burkhart}, \citenamefont {Pfaff}, \citenamefont {Zhang},
		\citenamefont {Chou}, \citenamefont {Campagne-Ibarcq}, \citenamefont
		{Reinhold}, \citenamefont {Frunzio}, \citenamefont {Girvin}, \citenamefont
		{Jiang} \emph {et~al.}}]{axline2018demand}%
	\BibitemOpen
	\bibfield  {author} {\bibinfo {author} {\bibfnamefont {Christopher~J}\
			\bibnamefont {Axline}}, \bibinfo {author} {\bibfnamefont {Luke~D}\
			\bibnamefont {Burkhart}}, \bibinfo {author} {\bibfnamefont {Wolfgang}\
			\bibnamefont {Pfaff}}, \bibinfo {author} {\bibfnamefont {Mengzhen}\
			\bibnamefont {Zhang}}, \bibinfo {author} {\bibfnamefont {Kevin}\ \bibnamefont
			{Chou}}, \bibinfo {author} {\bibfnamefont {Philippe}\ \bibnamefont
			{Campagne-Ibarcq}}, \bibinfo {author} {\bibfnamefont {Philip}\ \bibnamefont
			{Reinhold}}, \bibinfo {author} {\bibfnamefont {Luigi}\ \bibnamefont
			{Frunzio}}, \bibinfo {author} {\bibfnamefont {SM}~\bibnamefont {Girvin}},
		\bibinfo {author} {\bibfnamefont {Liang}\ \bibnamefont {Jiang}},  \emph
		{et~al.},\ }\bibfield  {title} {\enquote {\bibinfo {title} {On-demand quantum
				state transfer and entanglement between remote microwave cavity memories},}\
	}\href@noop {} {\bibfield  {journal} {\bibinfo  {journal} {Nature Physics}\
		}\textbf {\bibinfo {volume} {14}},\ \bibinfo {pages} {705--710} (\bibinfo
		{year} {2018})}\BibitemShut {NoStop}%
	\bibitem [{\citenamefont {Kurpiers}\ \emph {et~al.}(2018)\citenamefont
		{Kurpiers}, \citenamefont {Magnard}, \citenamefont {Walter}, \citenamefont
		{Royer}, \citenamefont {Pechal}, \citenamefont {Heinsoo}, \citenamefont
		{Salath{\'e}}, \citenamefont {Akin}, \citenamefont {Storz}, \citenamefont
		{Besse} \emph {et~al.}}]{kurpiers2018deterministic}%
	\BibitemOpen
	\bibfield  {author} {\bibinfo {author} {\bibfnamefont {Philipp}\ \bibnamefont
			{Kurpiers}}, \bibinfo {author} {\bibfnamefont {Paul}\ \bibnamefont
			{Magnard}}, \bibinfo {author} {\bibfnamefont {Theo}\ \bibnamefont {Walter}},
		\bibinfo {author} {\bibfnamefont {Baptiste}\ \bibnamefont {Royer}}, \bibinfo
		{author} {\bibfnamefont {Marek}\ \bibnamefont {Pechal}}, \bibinfo {author}
		{\bibfnamefont {Johannes}\ \bibnamefont {Heinsoo}}, \bibinfo {author}
		{\bibfnamefont {Yves}\ \bibnamefont {Salath{\'e}}}, \bibinfo {author}
		{\bibfnamefont {Abdulkadir}\ \bibnamefont {Akin}}, \bibinfo {author}
		{\bibfnamefont {Simon}\ \bibnamefont {Storz}}, \bibinfo {author}
		{\bibfnamefont {J-C}\ \bibnamefont {Besse}},  \emph {et~al.},\ }\bibfield
	{title} {\enquote {\bibinfo {title} {Deterministic quantum state transfer and
				remote entanglement using microwave photons},}\ }\href@noop {} {\bibfield
		{journal} {\bibinfo  {journal} {Nature}\ }\textbf {\bibinfo {volume} {558}},\
		\bibinfo {pages} {264--267} (\bibinfo {year} {2018})}\BibitemShut {NoStop}%
	\bibitem [{\citenamefont {Hermans}\ \emph {et~al.}(2022)\citenamefont
		{Hermans}, \citenamefont {Pompili}, \citenamefont {Beukers}, \citenamefont
		{Baier}, \citenamefont {Borregaard},\ and\ \citenamefont
		{Hanson}}]{hermans2022qubit}%
	\BibitemOpen
	\bibfield  {author} {\bibinfo {author} {\bibfnamefont {SLN}\ \bibnamefont
			{Hermans}}, \bibinfo {author} {\bibfnamefont {Matteo}\ \bibnamefont
			{Pompili}}, \bibinfo {author} {\bibfnamefont {HKC}\ \bibnamefont {Beukers}},
		\bibinfo {author} {\bibfnamefont {Simon}\ \bibnamefont {Baier}}, \bibinfo
		{author} {\bibfnamefont {Johannes}\ \bibnamefont {Borregaard}}, \ and\
		\bibinfo {author} {\bibfnamefont {Ronald}\ \bibnamefont {Hanson}},\
	}\bibfield  {title} {\enquote {\bibinfo {title} {Qubit teleportation between
				non-neighbouring nodes in a quantum network},}\ }\href@noop {} {\bibfield
		{journal} {\bibinfo  {journal} {Nature}\ }\textbf {\bibinfo {volume} {605}},\
		\bibinfo {pages} {663--668} (\bibinfo {year} {2022})}\BibitemShut {NoStop}%
	\bibitem [{\citenamefont {Wehner}\ \emph {et~al.}(2018)\citenamefont {Wehner},
		\citenamefont {Elkouss},\ and\ \citenamefont {Hanson}}]{wehner2018quantum}%
	\BibitemOpen
	\bibfield  {author} {\bibinfo {author} {\bibfnamefont {Stephanie}\
			\bibnamefont {Wehner}}, \bibinfo {author} {\bibfnamefont {David}\
			\bibnamefont {Elkouss}}, \ and\ \bibinfo {author} {\bibfnamefont {Ronald}\
			\bibnamefont {Hanson}},\ }\bibfield  {title} {\enquote {\bibinfo {title}
			{Quantum internet: A vision for the road ahead},}\ }\href@noop {} {\bibfield
		{journal} {\bibinfo  {journal} {Science}\ }\textbf {\bibinfo {volume}
			{362}},\ \bibinfo {pages} {eaam9288} (\bibinfo {year} {2018})}\BibitemShut
	{NoStop}%
	\bibitem [{\citenamefont {Awschalom}\ \emph {et~al.}(2021)\citenamefont
		{Awschalom}, \citenamefont {Berggren}, \citenamefont {Bernien}, \citenamefont
		{Bhave}, \citenamefont {Carr}, \citenamefont {Davids}, \citenamefont
		{Economou}, \citenamefont {Englund}, \citenamefont {Faraon}, \citenamefont
		{Fejer} \emph {et~al.}}]{awschalom2021development}%
	\BibitemOpen
	\bibfield  {author} {\bibinfo {author} {\bibfnamefont {David}\ \bibnamefont
			{Awschalom}}, \bibinfo {author} {\bibfnamefont {Karl~K}\ \bibnamefont
			{Berggren}}, \bibinfo {author} {\bibfnamefont {Hannes}\ \bibnamefont
			{Bernien}}, \bibinfo {author} {\bibfnamefont {Sunil}\ \bibnamefont {Bhave}},
		\bibinfo {author} {\bibfnamefont {Lincoln~D}\ \bibnamefont {Carr}}, \bibinfo
		{author} {\bibfnamefont {Paul}\ \bibnamefont {Davids}}, \bibinfo {author}
		{\bibfnamefont {Sophia~E}\ \bibnamefont {Economou}}, \bibinfo {author}
		{\bibfnamefont {Dirk}\ \bibnamefont {Englund}}, \bibinfo {author}
		{\bibfnamefont {Andrei}\ \bibnamefont {Faraon}}, \bibinfo {author}
		{\bibfnamefont {Martin}\ \bibnamefont {Fejer}},  \emph {et~al.},\ }\bibfield
	{title} {\enquote {\bibinfo {title} {Development of quantum interconnects
				(quics) for next-generation information technologies},}\ }\href@noop {}
	{\bibfield  {journal} {\bibinfo  {journal} {Prx Quantum}\ }\textbf {\bibinfo
			{volume} {2}},\ \bibinfo {pages} {017002} (\bibinfo {year}
		{2021})}\BibitemShut {NoStop}%
	\bibitem [{\citenamefont {Breuer}\ \emph {et~al.}(2016)\citenamefont {Breuer},
		\citenamefont {Laine}, \citenamefont {Piilo},\ and\ \citenamefont
		{Vacchini}}]{breuer2016colloquium}%
	\BibitemOpen
	\bibfield  {author} {\bibinfo {author} {\bibfnamefont {Heinz-Peter}\
			\bibnamefont {Breuer}}, \bibinfo {author} {\bibfnamefont {Elsi-Mari}\
			\bibnamefont {Laine}}, \bibinfo {author} {\bibfnamefont {Jyrki}\ \bibnamefont
			{Piilo}}, \ and\ \bibinfo {author} {\bibfnamefont {Bassano}\ \bibnamefont
			{Vacchini}},\ }\bibfield  {title} {\enquote {\bibinfo {title} {Colloquium:
				Non-markovian dynamics in open quantum systems},}\ }\href@noop {} {\bibfield
		{journal} {\bibinfo  {journal} {Reviews of Modern Physics}\ }\textbf
		{\bibinfo {volume} {88}},\ \bibinfo {pages} {021002} (\bibinfo {year}
		{2016})}\BibitemShut {NoStop}%
	\bibitem [{\citenamefont {He}\ and\ \citenamefont
		{Zhang}(2025)}]{he2025quantum}%
	\BibitemOpen
	\bibfield  {author} {\bibinfo {author} {\bibfnamefont {Yang}\ \bibnamefont
			{He}}\ and\ \bibinfo {author} {\bibfnamefont {Yu-Xiang}\ \bibnamefont
			{Zhang}},\ }\bibfield  {title} {\enquote {\bibinfo {title} {Quantum state
				transfer via a multimode resonator},}\ }\href@noop {} {\bibfield  {journal}
		{\bibinfo  {journal} {Physical Review Letters}\ }\textbf {\bibinfo {volume}
			{134}},\ \bibinfo {pages} {023602} (\bibinfo {year} {2025})}\BibitemShut
	{NoStop}%
	\bibitem [{\citenamefont {Chen}\ \emph {et~al.}(2014)\citenamefont {Chen},
		\citenamefont {Neill}, \citenamefont {Roushan}, \citenamefont {Leung},
		\citenamefont {Fang}, \citenamefont {Barends}, \citenamefont {Kelly},
		\citenamefont {Campbell}, \citenamefont {Chen}, \citenamefont {Chiaro} \emph
		{et~al.}}]{chen2014qubit}%
	\BibitemOpen
	\bibfield  {author} {\bibinfo {author} {\bibfnamefont {Yu}~\bibnamefont
			{Chen}}, \bibinfo {author} {\bibfnamefont {C}~\bibnamefont {Neill}}, \bibinfo
		{author} {\bibfnamefont {Pedram}\ \bibnamefont {Roushan}}, \bibinfo {author}
		{\bibfnamefont {Nelson}\ \bibnamefont {Leung}}, \bibinfo {author}
		{\bibfnamefont {Michael}\ \bibnamefont {Fang}}, \bibinfo {author}
		{\bibfnamefont {Rami}\ \bibnamefont {Barends}}, \bibinfo {author}
		{\bibfnamefont {Julian}\ \bibnamefont {Kelly}}, \bibinfo {author}
		{\bibfnamefont {Brooks}\ \bibnamefont {Campbell}}, \bibinfo {author}
		{\bibfnamefont {Z}~\bibnamefont {Chen}}, \bibinfo {author} {\bibfnamefont
			{Benjamin}\ \bibnamefont {Chiaro}},  \emph {et~al.},\ }\bibfield  {title}
	{\enquote {\bibinfo {title} {Qubit architecture with high coherence and fast
				tunable coupling},}\ }\href@noop {} {\bibfield  {journal} {\bibinfo
			{journal} {Physical review letters}\ }\textbf {\bibinfo {volume} {113}},\
		\bibinfo {pages} {220502} (\bibinfo {year} {2014})}\BibitemShut {NoStop}%
	\bibitem [{\citenamefont {Yan}\ \emph {et~al.}(2018)\citenamefont {Yan},
		\citenamefont {Krantz}, \citenamefont {Sung}, \citenamefont {Kjaergaard},
		\citenamefont {Campbell}, \citenamefont {Orlando}, \citenamefont
		{Gustavsson},\ and\ \citenamefont {Oliver}}]{yan2018tunable}%
	\BibitemOpen
	\bibfield  {author} {\bibinfo {author} {\bibfnamefont {Fei}\ \bibnamefont
			{Yan}}, \bibinfo {author} {\bibfnamefont {Philip}\ \bibnamefont {Krantz}},
		\bibinfo {author} {\bibfnamefont {Youngkyu}\ \bibnamefont {Sung}}, \bibinfo
		{author} {\bibfnamefont {Morten}\ \bibnamefont {Kjaergaard}}, \bibinfo
		{author} {\bibfnamefont {Daniel~L}\ \bibnamefont {Campbell}}, \bibinfo
		{author} {\bibfnamefont {Terry~P}\ \bibnamefont {Orlando}}, \bibinfo {author}
		{\bibfnamefont {Simon}\ \bibnamefont {Gustavsson}}, \ and\ \bibinfo {author}
		{\bibfnamefont {William~D}\ \bibnamefont {Oliver}},\ }\bibfield  {title}
	{\enquote {\bibinfo {title} {Tunable coupling scheme for implementing
				high-fidelity two-qubit gates},}\ }\href@noop {} {\bibfield  {journal}
		{\bibinfo  {journal} {Physical Review Applied}\ }\textbf {\bibinfo {volume}
			{10}},\ \bibinfo {pages} {054062} (\bibinfo {year} {2018})}\BibitemShut
	{NoStop}%
	\bibitem [{\citenamefont {Vitanov}\ \emph {et~al.}(2017)\citenamefont
		{Vitanov}, \citenamefont {Rangelov}, \citenamefont {Shore},\ and\
		\citenamefont {Bergmann}}]{vitanov2017stimulated}%
	\BibitemOpen
	\bibfield  {author} {\bibinfo {author} {\bibfnamefont {Nikolay~V}\
			\bibnamefont {Vitanov}}, \bibinfo {author} {\bibfnamefont {Andon~A}\
			\bibnamefont {Rangelov}}, \bibinfo {author} {\bibfnamefont {Bruce~W}\
			\bibnamefont {Shore}}, \ and\ \bibinfo {author} {\bibfnamefont {Klaas}\
			\bibnamefont {Bergmann}},\ }\bibfield  {title} {\enquote {\bibinfo {title}
			{Stimulated raman adiabatic passage in physics, chemistry, and beyond},}\
	}\href@noop {} {\bibfield  {journal} {\bibinfo  {journal} {Reviews of Modern
				Physics}\ }\textbf {\bibinfo {volume} {89}},\ \bibinfo {pages} {015006}
		(\bibinfo {year} {2017})}\BibitemShut {NoStop}%
	\bibitem [{\citenamefont {Chang}\ \emph {et~al.}(2020)\citenamefont {Chang},
		\citenamefont {Zhong}, \citenamefont {Bienfait}, \citenamefont {Chou},
		\citenamefont {Conner}, \citenamefont {Dumur}, \citenamefont {Grebel},
		\citenamefont {Peairs}, \citenamefont {Povey}, \citenamefont {Satzinger}
		\emph {et~al.}}]{chang2020remote}%
	\BibitemOpen
	\bibfield  {author} {\bibinfo {author} {\bibfnamefont {H-S}\ \bibnamefont
			{Chang}}, \bibinfo {author} {\bibfnamefont {YP}~\bibnamefont {Zhong}},
		\bibinfo {author} {\bibfnamefont {Audrey}\ \bibnamefont {Bienfait}}, \bibinfo
		{author} {\bibfnamefont {M-H}\ \bibnamefont {Chou}}, \bibinfo {author}
		{\bibfnamefont {Christopher~R}\ \bibnamefont {Conner}}, \bibinfo {author}
		{\bibfnamefont {{\'E}tienne}\ \bibnamefont {Dumur}}, \bibinfo {author}
		{\bibfnamefont {Joel}\ \bibnamefont {Grebel}}, \bibinfo {author}
		{\bibfnamefont {Gregory~A}\ \bibnamefont {Peairs}}, \bibinfo {author}
		{\bibfnamefont {Rhys~G}\ \bibnamefont {Povey}}, \bibinfo {author}
		{\bibfnamefont {Kevin~J}\ \bibnamefont {Satzinger}},  \emph {et~al.},\
	}\bibfield  {title} {\enquote {\bibinfo {title} {Remote entanglement via
				adiabatic passage using a tunably dissipative quantum communication
				system},}\ }\href@noop {} {\bibfield  {journal} {\bibinfo  {journal}
			{Physical Review Letters}\ }\textbf {\bibinfo {volume} {124}},\ \bibinfo
		{pages} {240502} (\bibinfo {year} {2020})}\BibitemShut {NoStop}%
	\bibitem [{\citenamefont {Sundaresan}\ \emph {et~al.}(2015)\citenamefont
		{Sundaresan}, \citenamefont {Liu}, \citenamefont {Sadri}, \citenamefont
		{Sz{\H{o}}cs}, \citenamefont {Underwood}, \citenamefont {Malekakhlagh},
		\citenamefont {T{\"u}reci},\ and\ \citenamefont
		{Houck}}]{sundaresan2015beyond}%
	\BibitemOpen
	\bibfield  {author} {\bibinfo {author} {\bibfnamefont {Neereja~M}\
			\bibnamefont {Sundaresan}}, \bibinfo {author} {\bibfnamefont {Yanbing}\
			\bibnamefont {Liu}}, \bibinfo {author} {\bibfnamefont {Darius}\ \bibnamefont
			{Sadri}}, \bibinfo {author} {\bibfnamefont {L{\'a}szl{\'o}~J}\ \bibnamefont
			{Sz{\H{o}}cs}}, \bibinfo {author} {\bibfnamefont {Devin~L}\ \bibnamefont
			{Underwood}}, \bibinfo {author} {\bibfnamefont {Moein}\ \bibnamefont
			{Malekakhlagh}}, \bibinfo {author} {\bibfnamefont {Hakan~E}\ \bibnamefont
			{T{\"u}reci}}, \ and\ \bibinfo {author} {\bibfnamefont {Andrew~A}\
			\bibnamefont {Houck}},\ }\bibfield  {title} {\enquote {\bibinfo {title}
			{Beyond strong coupling in a multimode cavity},}\ }\href@noop {} {\bibfield
		{journal} {\bibinfo  {journal} {Physical Review X}\ }\textbf {\bibinfo
			{volume} {5}},\ \bibinfo {pages} {021035} (\bibinfo {year}
		{2015})}\BibitemShut {NoStop}%
	\bibitem [{\citenamefont {Kuzmin}\ \emph {et~al.}(2019)\citenamefont {Kuzmin},
		\citenamefont {Mehta}, \citenamefont {Grabon}, \citenamefont {Mencia},\ and\
		\citenamefont {Manucharyan}}]{kuzmin2019superstrong}%
	\BibitemOpen
	\bibfield  {author} {\bibinfo {author} {\bibfnamefont {Roman}\ \bibnamefont
			{Kuzmin}}, \bibinfo {author} {\bibfnamefont {Nitish}\ \bibnamefont {Mehta}},
		\bibinfo {author} {\bibfnamefont {Nicholas}\ \bibnamefont {Grabon}}, \bibinfo
		{author} {\bibfnamefont {Raymond}\ \bibnamefont {Mencia}}, \ and\ \bibinfo
		{author} {\bibfnamefont {Vladimir~E}\ \bibnamefont {Manucharyan}},\
	}\bibfield  {title} {\enquote {\bibinfo {title} {Superstrong coupling in
				circuit quantum electrodynamics},}\ }\href@noop {} {\bibfield  {journal}
		{\bibinfo  {journal} {npj Quantum Information}\ }\textbf {\bibinfo {volume}
			{5}},\ \bibinfo {pages} {20} (\bibinfo {year} {2019})}\BibitemShut {NoStop}%
	\bibitem [{\citenamefont {Satzinger}\ \emph {et~al.}(2018)\citenamefont
		{Satzinger}, \citenamefont {Zhong}, \citenamefont {Chang}, \citenamefont
		{Peairs}, \citenamefont {Bienfait}, \citenamefont {Chou}, \citenamefont
		{Cleland}, \citenamefont {Conner}, \citenamefont {Dumur}, \citenamefont
		{Grebel} \emph {et~al.}}]{satzinger2018quantum}%
	\BibitemOpen
	\bibfield  {author} {\bibinfo {author} {\bibfnamefont {Kevin~Joseph}\
			\bibnamefont {Satzinger}}, \bibinfo {author} {\bibfnamefont {YP}~\bibnamefont
			{Zhong}}, \bibinfo {author} {\bibfnamefont {H-S}\ \bibnamefont {Chang}},
		\bibinfo {author} {\bibfnamefont {Gregory~A}\ \bibnamefont {Peairs}},
		\bibinfo {author} {\bibfnamefont {Audrey}\ \bibnamefont {Bienfait}}, \bibinfo
		{author} {\bibfnamefont {Ming-Han}\ \bibnamefont {Chou}}, \bibinfo {author}
		{\bibfnamefont {AY}~\bibnamefont {Cleland}}, \bibinfo {author} {\bibfnamefont
			{Cristopher~R}\ \bibnamefont {Conner}}, \bibinfo {author} {\bibfnamefont
			{{\'E}tienne}\ \bibnamefont {Dumur}}, \bibinfo {author} {\bibfnamefont
			{Joel}\ \bibnamefont {Grebel}},  \emph {et~al.},\ }\bibfield  {title}
	{\enquote {\bibinfo {title} {Quantum control of surface acoustic-wave
				phonons},}\ }\href@noop {} {\bibfield  {journal} {\bibinfo  {journal}
			{Nature}\ }\textbf {\bibinfo {volume} {563}},\ \bibinfo {pages} {661--665}
		(\bibinfo {year} {2018})}\BibitemShut {NoStop}%
	\bibitem [{\citenamefont {Moores}\ \emph {et~al.}(2018)\citenamefont {Moores},
		\citenamefont {Sletten}, \citenamefont {Viennot},\ and\ \citenamefont
		{Lehnert}}]{moores2018cavity}%
	\BibitemOpen
	\bibfield  {author} {\bibinfo {author} {\bibfnamefont {Bradley~A}\
			\bibnamefont {Moores}}, \bibinfo {author} {\bibfnamefont {Lucas~R}\
			\bibnamefont {Sletten}}, \bibinfo {author} {\bibfnamefont {Jeremie~J}\
			\bibnamefont {Viennot}}, \ and\ \bibinfo {author} {\bibfnamefont
			{KW}~\bibnamefont {Lehnert}},\ }\bibfield  {title} {\enquote {\bibinfo
			{title} {Cavity quantum acoustic device in the multimode strong coupling
				regime},}\ }\href@noop {} {\bibfield  {journal} {\bibinfo  {journal}
			{Physical review letters}\ }\textbf {\bibinfo {volume} {120}},\ \bibinfo
		{pages} {227701} (\bibinfo {year} {2018})}\BibitemShut {NoStop}%
	\bibitem [{\citenamefont {Scigliuzzo}\ \emph {et~al.}(2025)\citenamefont
		{Scigliuzzo}, \citenamefont {Peyruchat}, \citenamefont {Marabini},
		\citenamefont {Becker}, \citenamefont {Jouanny}, \citenamefont {Delsing},\
		and\ \citenamefont {Scarlino}}]{scigliuzzo2025quantum}%
	\BibitemOpen
	\bibfield  {author} {\bibinfo {author} {\bibfnamefont {Marco}\ \bibnamefont
			{Scigliuzzo}}, \bibinfo {author} {\bibfnamefont {L{\'e}o}\ \bibnamefont
			{Peyruchat}}, \bibinfo {author} {\bibfnamefont {Riccardo~Maria}\ \bibnamefont
			{Marabini}}, \bibinfo {author} {\bibfnamefont {Carla}\ \bibnamefont
			{Becker}}, \bibinfo {author} {\bibfnamefont {Vincent}\ \bibnamefont
			{Jouanny}}, \bibinfo {author} {\bibfnamefont {Per}\ \bibnamefont {Delsing}},
		\ and\ \bibinfo {author} {\bibfnamefont {Pasquale}\ \bibnamefont
			{Scarlino}},\ }\bibfield  {title} {\enquote {\bibinfo {title} {Quantum
				acoustics with tunable nonlinearity in the superstrong coupling regime},}\
	}\href@noop {} {\bibfield  {journal} {\bibinfo  {journal} {arXiv preprint
				arXiv:2505.24865}\ } (\bibinfo {year} {2025})}\BibitemShut {NoStop}%
	\bibitem [{\citenamefont {Cirac}\ \emph {et~al.}(1997)\citenamefont {Cirac},
		\citenamefont {Zoller}, \citenamefont {Kimble},\ and\ \citenamefont
		{Mabuchi}}]{cirac1997quantum}%
	\BibitemOpen
	\bibfield  {author} {\bibinfo {author} {\bibfnamefont {Juan~Ignacio}\
			\bibnamefont {Cirac}}, \bibinfo {author} {\bibfnamefont {Peter}\ \bibnamefont
			{Zoller}}, \bibinfo {author} {\bibfnamefont {H~Jeff}\ \bibnamefont {Kimble}},
		\ and\ \bibinfo {author} {\bibfnamefont {Hideo}\ \bibnamefont {Mabuchi}},\
	}\bibfield  {title} {\enquote {\bibinfo {title} {Quantum state transfer and
				entanglement distribution among distant nodes in a quantum network},}\
	}\href@noop {} {\bibfield  {journal} {\bibinfo  {journal} {Physical Review
				Letters}\ }\textbf {\bibinfo {volume} {78}},\ \bibinfo {pages} {3221}
		(\bibinfo {year} {1997})}\BibitemShut {NoStop}%
	\bibitem [{\citenamefont {Korotkov}(2011)}]{korotkov2011flying}%
	\BibitemOpen
	\bibfield  {author} {\bibinfo {author} {\bibfnamefont {Alexander~N}\
			\bibnamefont {Korotkov}},\ }\bibfield  {title} {\enquote {\bibinfo {title}
			{Flying microwave qubits with nearly perfect transfer efficiency},}\
	}\href@noop {} {\bibfield  {journal} {\bibinfo  {journal} {Physical Review
				B—Condensed Matter and Materials Physics}\ }\textbf {\bibinfo {volume}
			{84}},\ \bibinfo {pages} {014510} (\bibinfo {year} {2011})}\BibitemShut
	{NoStop}%
	\bibitem [{\citenamefont {Sete}\ \emph {et~al.}(2015)\citenamefont {Sete},
		\citenamefont {Mlinar},\ and\ \citenamefont {Korotkov}}]{sete2015robust}%
	\BibitemOpen
	\bibfield  {author} {\bibinfo {author} {\bibfnamefont {Eyob~A}\ \bibnamefont
			{Sete}}, \bibinfo {author} {\bibfnamefont {Eric}\ \bibnamefont {Mlinar}}, \
		and\ \bibinfo {author} {\bibfnamefont {Alexander~N}\ \bibnamefont
			{Korotkov}},\ }\bibfield  {title} {\enquote {\bibinfo {title} {Robust quantum
				state transfer using tunable couplers},}\ }\href@noop {} {\bibfield
		{journal} {\bibinfo  {journal} {Physical Review B}\ }\textbf {\bibinfo
			{volume} {91}},\ \bibinfo {pages} {144509} (\bibinfo {year}
		{2015})}\BibitemShut {NoStop}%
	\bibitem [{\citenamefont {Bienfait}\ \emph {et~al.}(2019)\citenamefont
		{Bienfait}, \citenamefont {Satzinger}, \citenamefont {Zhong}, \citenamefont
		{Chang}, \citenamefont {Chou}, \citenamefont {Conner}, \citenamefont {Dumur},
		\citenamefont {Grebel}, \citenamefont {Peairs}, \citenamefont {Povey} \emph
		{et~al.}}]{bienfait2019phonon}%
	\BibitemOpen
	\bibfield  {author} {\bibinfo {author} {\bibfnamefont {Audrey}\ \bibnamefont
			{Bienfait}}, \bibinfo {author} {\bibfnamefont {Kevin~J}\ \bibnamefont
			{Satzinger}}, \bibinfo {author} {\bibfnamefont {YP}~\bibnamefont {Zhong}},
		\bibinfo {author} {\bibfnamefont {H-S}\ \bibnamefont {Chang}}, \bibinfo
		{author} {\bibfnamefont {M-H}\ \bibnamefont {Chou}}, \bibinfo {author}
		{\bibfnamefont {Chris~R}\ \bibnamefont {Conner}}, \bibinfo {author}
		{\bibfnamefont {{\'E}}~\bibnamefont {Dumur}}, \bibinfo {author}
		{\bibfnamefont {Joel}\ \bibnamefont {Grebel}}, \bibinfo {author}
		{\bibfnamefont {Gregory~A}\ \bibnamefont {Peairs}}, \bibinfo {author}
		{\bibfnamefont {Rhys~G}\ \bibnamefont {Povey}},  \emph {et~al.},\ }\bibfield
	{title} {\enquote {\bibinfo {title} {Phonon-mediated quantum state transfer
				and remote qubit entanglement},}\ }\href@noop {} {\bibfield  {journal}
		{\bibinfo  {journal} {Science}\ }\textbf {\bibinfo {volume} {364}},\ \bibinfo
		{pages} {368--371} (\bibinfo {year} {2019})}\BibitemShut {NoStop}%
	\bibitem [{\citenamefont {Qiao}\ \emph {et~al.}(2023)\citenamefont {Qiao},
		\citenamefont {Dumur}, \citenamefont {Andersson}, \citenamefont {Yan},
		\citenamefont {Chou}, \citenamefont {Grebel}, \citenamefont {Conner},
		\citenamefont {Joshi}, \citenamefont {Miller}, \citenamefont {Povey} \emph
		{et~al.}}]{qiao2023splitting}%
	\BibitemOpen
	\bibfield  {author} {\bibinfo {author} {\bibfnamefont {Hong}\ \bibnamefont
			{Qiao}}, \bibinfo {author} {\bibfnamefont {{\'E}tienne}\ \bibnamefont
			{Dumur}}, \bibinfo {author} {\bibfnamefont {Gustav}\ \bibnamefont
			{Andersson}}, \bibinfo {author} {\bibfnamefont {Haoxiong}\ \bibnamefont
			{Yan}}, \bibinfo {author} {\bibfnamefont {M-H}\ \bibnamefont {Chou}},
		\bibinfo {author} {\bibfnamefont {Joel}\ \bibnamefont {Grebel}}, \bibinfo
		{author} {\bibfnamefont {Christopher~R}\ \bibnamefont {Conner}}, \bibinfo
		{author} {\bibfnamefont {Yash~J}\ \bibnamefont {Joshi}}, \bibinfo {author}
		{\bibfnamefont {Jacob~M}\ \bibnamefont {Miller}}, \bibinfo {author}
		{\bibfnamefont {Rhys~G}\ \bibnamefont {Povey}},  \emph {et~al.},\ }\bibfield
	{title} {\enquote {\bibinfo {title} {Splitting phonons: Building a platform
				for linear mechanical quantum computing},}\ }\href@noop {} {\bibfield
		{journal} {\bibinfo  {journal} {Science}\ }\textbf {\bibinfo {volume}
			{380}},\ \bibinfo {pages} {1030--1033} (\bibinfo {year} {2023})}\BibitemShut
	{NoStop}%
	\bibitem [{\citenamefont {Grebel}\ \emph {et~al.}(2024)\citenamefont {Grebel},
		\citenamefont {Yan}, \citenamefont {Chou}, \citenamefont {Andersson},
		\citenamefont {Conner}, \citenamefont {Joshi}, \citenamefont {Miller},
		\citenamefont {Povey}, \citenamefont {Qiao}, \citenamefont {Wu} \emph
		{et~al.}}]{grebel2024bidirectional}%
	\BibitemOpen
	\bibfield  {author} {\bibinfo {author} {\bibfnamefont {Joel}\ \bibnamefont
			{Grebel}}, \bibinfo {author} {\bibfnamefont {Haoxiong}\ \bibnamefont {Yan}},
		\bibinfo {author} {\bibfnamefont {Ming-Han}\ \bibnamefont {Chou}}, \bibinfo
		{author} {\bibfnamefont {Gustav}\ \bibnamefont {Andersson}}, \bibinfo
		{author} {\bibfnamefont {Christopher~R}\ \bibnamefont {Conner}}, \bibinfo
		{author} {\bibfnamefont {Yash~J}\ \bibnamefont {Joshi}}, \bibinfo {author}
		{\bibfnamefont {Jacob~M}\ \bibnamefont {Miller}}, \bibinfo {author}
		{\bibfnamefont {Rhys~G}\ \bibnamefont {Povey}}, \bibinfo {author}
		{\bibfnamefont {Hong}\ \bibnamefont {Qiao}}, \bibinfo {author} {\bibfnamefont
			{Xuntao}\ \bibnamefont {Wu}},  \emph {et~al.},\ }\bibfield  {title} {\enquote
		{\bibinfo {title} {Bidirectional multiphoton communication between remote
				superconducting nodes},}\ }\href@noop {} {\bibfield  {journal} {\bibinfo
			{journal} {Physical Review Letters}\ }\textbf {\bibinfo {volume} {132}},\
		\bibinfo {pages} {047001} (\bibinfo {year} {2024})}\BibitemShut {NoStop}%
	\bibitem [{\citenamefont {Vogell}\ \emph {et~al.}(2017)\citenamefont {Vogell},
		\citenamefont {Vermersch}, \citenamefont {Northup}, \citenamefont {Lanyon},\
		and\ \citenamefont {Muschik}}]{vogell2017deterministic}%
	\BibitemOpen
	\bibfield  {author} {\bibinfo {author} {\bibfnamefont {B}~\bibnamefont
			{Vogell}}, \bibinfo {author} {\bibfnamefont {B}~\bibnamefont {Vermersch}},
		\bibinfo {author} {\bibfnamefont {TE}~\bibnamefont {Northup}}, \bibinfo
		{author} {\bibfnamefont {BP}~\bibnamefont {Lanyon}}, \ and\ \bibinfo {author}
		{\bibfnamefont {CA}~\bibnamefont {Muschik}},\ }\bibfield  {title} {\enquote
		{\bibinfo {title} {Deterministic quantum state transfer between remote qubits
				in cavities},}\ }\href@noop {} {\bibfield  {journal} {\bibinfo  {journal}
			{Quantum Science and Technology}\ }\textbf {\bibinfo {volume} {2}},\ \bibinfo
		{pages} {045003} (\bibinfo {year} {2017})}\BibitemShut {NoStop}%
	\bibitem [{\citenamefont {Malekakhlagh}\ \emph {et~al.}(2024)\citenamefont
		{Malekakhlagh}, \citenamefont {Phung}, \citenamefont {Puzzuoli},
		\citenamefont {Heya}, \citenamefont {Sundaresan},\ and\ \citenamefont
		{Orcutt}}]{malekakhlagh2024enhanced}%
	\BibitemOpen
	\bibfield  {author} {\bibinfo {author} {\bibfnamefont {Moein}\ \bibnamefont
			{Malekakhlagh}}, \bibinfo {author} {\bibfnamefont {Timothy}\ \bibnamefont
			{Phung}}, \bibinfo {author} {\bibfnamefont {Daniel}\ \bibnamefont
			{Puzzuoli}}, \bibinfo {author} {\bibfnamefont {Kentaro}\ \bibnamefont
			{Heya}}, \bibinfo {author} {\bibfnamefont {Neereja}\ \bibnamefont
			{Sundaresan}}, \ and\ \bibinfo {author} {\bibfnamefont {Jason}\ \bibnamefont
			{Orcutt}},\ }\bibfield  {title} {\enquote {\bibinfo {title} {Enhanced quantum
				state transfer and bell-state generation over long-range multimode
				interconnects via superadiabatic transitionless driving},}\ }\href@noop {}
	{\bibfield  {journal} {\bibinfo  {journal} {Physical Review Applied}\
		}\textbf {\bibinfo {volume} {22}},\ \bibinfo {pages} {024006} (\bibinfo
		{year} {2024})}\BibitemShut {NoStop}%
	\bibitem [{\citenamefont {Calaj{\'o}}\ \emph {et~al.}(2016)\citenamefont
		{Calaj{\'o}}, \citenamefont {Ciccarello}, \citenamefont {Chang},\ and\
		\citenamefont {Rabl}}]{calajo2016atom}%
	\BibitemOpen
	\bibfield  {author} {\bibinfo {author} {\bibfnamefont {Giuseppe}\
			\bibnamefont {Calaj{\'o}}}, \bibinfo {author} {\bibfnamefont {Francesco}\
			\bibnamefont {Ciccarello}}, \bibinfo {author} {\bibfnamefont {Darrick}\
			\bibnamefont {Chang}}, \ and\ \bibinfo {author} {\bibfnamefont {Peter}\
			\bibnamefont {Rabl}},\ }\bibfield  {title} {\enquote {\bibinfo {title}
			{Atom-field dressed states in slow-light waveguide qed},}\ }\href@noop {}
	{\bibfield  {journal} {\bibinfo  {journal} {Physical Review A}\ }\textbf
		{\bibinfo {volume} {93}},\ \bibinfo {pages} {033833} (\bibinfo {year}
		{2016})}\BibitemShut {NoStop}%
	\bibitem [{\citenamefont {Milonni}\ \emph {et~al.}(1983)\citenamefont
		{Milonni}, \citenamefont {Ackerhalt}, \citenamefont {Galbraith},\ and\
		\citenamefont {Shih}}]{milonni1983exponential}%
	\BibitemOpen
	\bibfield  {author} {\bibinfo {author} {\bibfnamefont {PW}~\bibnamefont
			{Milonni}}, \bibinfo {author} {\bibfnamefont {JR}~\bibnamefont {Ackerhalt}},
		\bibinfo {author} {\bibfnamefont {HW}~\bibnamefont {Galbraith}}, \ and\
		\bibinfo {author} {\bibfnamefont {Mei-Li}\ \bibnamefont {Shih}},\ }\bibfield
	{title} {\enquote {\bibinfo {title} {Exponential decay, recurrences, and
				quantum-mechanical spreading in a quasicontinuum model},}\ }\href@noop {}
	{\bibfield  {journal} {\bibinfo  {journal} {Physical Review A}\ }\textbf
		{\bibinfo {volume} {28}},\ \bibinfo {pages} {32} (\bibinfo {year}
		{1983})}\BibitemShut {NoStop}%
	\bibitem [{\citenamefont {Krimer}\ \emph {et~al.}(2014)\citenamefont {Krimer},
		\citenamefont {Liertzer}, \citenamefont {Rotter},\ and\ \citenamefont
		{T{\"u}reci}}]{krimer2014route}%
	\BibitemOpen
	\bibfield  {author} {\bibinfo {author} {\bibfnamefont {Dmitry~O}\
			\bibnamefont {Krimer}}, \bibinfo {author} {\bibfnamefont {Matthias}\
			\bibnamefont {Liertzer}}, \bibinfo {author} {\bibfnamefont {Stefan}\
			\bibnamefont {Rotter}}, \ and\ \bibinfo {author} {\bibfnamefont {Hakan~E}\
			\bibnamefont {T{\"u}reci}},\ }\bibfield  {title} {\enquote {\bibinfo {title}
			{Route from spontaneous decay to complex multimode dynamics in cavity qed},}\
	}\href@noop {} {\bibfield  {journal} {\bibinfo  {journal} {Physical Review
				A}\ }\textbf {\bibinfo {volume} {89}},\ \bibinfo {pages} {033820} (\bibinfo
		{year} {2014})}\BibitemShut {NoStop}%
	\bibitem [{\citenamefont {Meiser}\ and\ \citenamefont
		{Meystre}(2006)}]{meiser2006superstrong}%
	\BibitemOpen
	\bibfield  {author} {\bibinfo {author} {\bibfnamefont {D}~\bibnamefont
			{Meiser}}\ and\ \bibinfo {author} {\bibfnamefont {P}~\bibnamefont
			{Meystre}},\ }\bibfield  {title} {\enquote {\bibinfo {title} {Superstrong
				coupling regime of cavity quantum electrodynamics},}\ }\href@noop {}
	{\bibfield  {journal} {\bibinfo  {journal} {Physical Review A—Atomic,
				Molecular, and Optical Physics}\ }\textbf {\bibinfo {volume} {74}},\ \bibinfo
		{pages} {065801} (\bibinfo {year} {2006})}\BibitemShut {NoStop}%
	\bibitem [{\citenamefont {Zhong}\ \emph {et~al.}(2019)\citenamefont {Zhong},
		\citenamefont {Chang}, \citenamefont {Satzinger}, \citenamefont {Chou},
		\citenamefont {Bienfait}, \citenamefont {Conner}, \citenamefont {Dumur},
		\citenamefont {Grebel}, \citenamefont {Peairs}, \citenamefont {Povey} \emph
		{et~al.}}]{zhong2019violating}%
	\BibitemOpen
	\bibfield  {author} {\bibinfo {author} {\bibfnamefont {YP}~\bibnamefont
			{Zhong}}, \bibinfo {author} {\bibfnamefont {H-S}\ \bibnamefont {Chang}},
		\bibinfo {author} {\bibfnamefont {KJ}~\bibnamefont {Satzinger}}, \bibinfo
		{author} {\bibfnamefont {M-H}\ \bibnamefont {Chou}}, \bibinfo {author}
		{\bibfnamefont {Audrey}\ \bibnamefont {Bienfait}}, \bibinfo {author}
		{\bibfnamefont {CR}~\bibnamefont {Conner}}, \bibinfo {author} {\bibfnamefont
			{{\'E}}~\bibnamefont {Dumur}}, \bibinfo {author} {\bibfnamefont {Joel}\
			\bibnamefont {Grebel}}, \bibinfo {author} {\bibfnamefont {GA}~\bibnamefont
			{Peairs}}, \bibinfo {author} {\bibfnamefont {RG}~\bibnamefont {Povey}},
		\emph {et~al.},\ }\bibfield  {title} {\enquote {\bibinfo {title} {Violating
				bell’s inequality with remotely connected superconducting qubits},}\
	}\href@noop {} {\bibfield  {journal} {\bibinfo  {journal} {Nature Physics}\
		}\textbf {\bibinfo {volume} {15}},\ \bibinfo {pages} {741--744} (\bibinfo
		{year} {2019})}\BibitemShut {NoStop}%
	\bibitem [{\citenamefont {Kjaergaard}\ \emph {et~al.}(2020)\citenamefont
		{Kjaergaard}, \citenamefont {Schwartz}, \citenamefont {Braum{\"u}ller},
		\citenamefont {Krantz}, \citenamefont {Wang}, \citenamefont {Gustavsson},\
		and\ \citenamefont {Oliver}}]{kjaergaard2020superconducting}%
	\BibitemOpen
	\bibfield  {author} {\bibinfo {author} {\bibfnamefont {Morten}\ \bibnamefont
			{Kjaergaard}}, \bibinfo {author} {\bibfnamefont {Mollie~E}\ \bibnamefont
			{Schwartz}}, \bibinfo {author} {\bibfnamefont {Jochen}\ \bibnamefont
			{Braum{\"u}ller}}, \bibinfo {author} {\bibfnamefont {Philip}\ \bibnamefont
			{Krantz}}, \bibinfo {author} {\bibfnamefont {Joel I-J}\ \bibnamefont {Wang}},
		\bibinfo {author} {\bibfnamefont {Simon}\ \bibnamefont {Gustavsson}}, \ and\
		\bibinfo {author} {\bibfnamefont {William~D}\ \bibnamefont {Oliver}},\
	}\bibfield  {title} {\enquote {\bibinfo {title} {Superconducting qubits:
				Current state of play},}\ }\href@noop {} {\bibfield  {journal} {\bibinfo
			{journal} {Annual Review of Condensed Matter Physics}\ }\textbf {\bibinfo
			{volume} {11}},\ \bibinfo {pages} {369--395} (\bibinfo {year}
		{2020})}\BibitemShut {NoStop}%
	\bibitem [{\citenamefont {Blais}\ \emph {et~al.}(2021)\citenamefont {Blais},
		\citenamefont {Grimsmo}, \citenamefont {Girvin},\ and\ \citenamefont
		{Wallraff}}]{blais2021circuit}%
	\BibitemOpen
	\bibfield  {author} {\bibinfo {author} {\bibfnamefont {Alexandre}\
			\bibnamefont {Blais}}, \bibinfo {author} {\bibfnamefont {Arne~L}\
			\bibnamefont {Grimsmo}}, \bibinfo {author} {\bibfnamefont {Steven~M}\
			\bibnamefont {Girvin}}, \ and\ \bibinfo {author} {\bibfnamefont {Andreas}\
			\bibnamefont {Wallraff}},\ }\bibfield  {title} {\enquote {\bibinfo {title}
			{Circuit quantum electrodynamics},}\ }\href@noop {} {\bibfield  {journal}
		{\bibinfo  {journal} {Reviews of Modern Physics}\ }\textbf {\bibinfo {volume}
			{93}},\ \bibinfo {pages} {025005} (\bibinfo {year} {2021})}\BibitemShut
	{NoStop}%
	\bibitem [{\citenamefont {Zhang}\ \emph {et~al.}(2015)\citenamefont {Zhang},
		\citenamefont {Zou}, \citenamefont {Jiang},\ and\ \citenamefont
		{Tang}}]{zhang2017multimode}%
	\BibitemOpen
	\bibfield  {author} {\bibinfo {author} {\bibfnamefont {X.}~\bibnamefont
			{Zhang}}, \bibinfo {author} {\bibfnamefont {C.-L.}\ \bibnamefont {Zou}},
		\bibinfo {author} {\bibfnamefont {L.}~\bibnamefont {Jiang}}, \ and\ \bibinfo
		{author} {\bibfnamefont {H.~X.}\ \bibnamefont {Tang}},\ }\bibfield  {title}
	{\enquote {\bibinfo {title} {Multimode circuit quantum electrodynamics with
				hybrid metamaterial transmission lines},}\ }\href@noop {} {\bibfield
		{journal} {\bibinfo  {journal} {Nature Communications}\ }\textbf {\bibinfo
			{volume} {6}},\ \bibinfo {pages} {8914} (\bibinfo {year} {2015})}\BibitemShut
	{NoStop}%
	\bibitem [{\citenamefont {Zivari}\ \emph {et~al.}(2022)\citenamefont {Zivari},
		\citenamefont {Fiaschi}, \citenamefont {Burgwal}, \citenamefont {Verhagen},
		\citenamefont {Stockill},\ and\ \citenamefont
		{Gr{\"o}blacher}}]{zivari2022chip}%
	\BibitemOpen
	\bibfield  {author} {\bibinfo {author} {\bibfnamefont {Amirparsa}\
			\bibnamefont {Zivari}}, \bibinfo {author} {\bibfnamefont {Niccol{\`o}}\
			\bibnamefont {Fiaschi}}, \bibinfo {author} {\bibfnamefont {Roel}\
			\bibnamefont {Burgwal}}, \bibinfo {author} {\bibfnamefont {Ewold}\
			\bibnamefont {Verhagen}}, \bibinfo {author} {\bibfnamefont {Robert}\
			\bibnamefont {Stockill}}, \ and\ \bibinfo {author} {\bibfnamefont {Simon}\
			\bibnamefont {Gr{\"o}blacher}},\ }\bibfield  {title} {\enquote {\bibinfo
			{title} {On-chip distribution of quantum information using traveling
				phonons},}\ }\href@noop {} {\bibfield  {journal} {\bibinfo  {journal}
			{Science advances}\ }\textbf {\bibinfo {volume} {8}},\ \bibinfo {pages}
		{eadd2811} (\bibinfo {year} {2022})}\BibitemShut {NoStop}%
	\bibitem [{\citenamefont {Chen}\ \emph {et~al.}(2016)\citenamefont {Chen},
		\citenamefont {Kelly}, \citenamefont {Quintana}, \citenamefont {Barends},
		\citenamefont {Campbell}, \citenamefont {Chen}, \citenamefont {Chiaro},
		\citenamefont {Dunsworth}, \citenamefont {Fowler}, \citenamefont {Jeffrey},
		\citenamefont {Megrant}, \citenamefont {Mutus}, \citenamefont {Neill},
		\citenamefont {O'Malley}, \citenamefont {Roushan}, \citenamefont {Sank},
		\citenamefont {Vainsencher}, \citenamefont {Wenner}, \citenamefont {White},
		\citenamefont {Cleland},\ and\ \citenamefont {Martinis}}]{chen2016measuring}%
	\BibitemOpen
	\bibfield  {author} {\bibinfo {author} {\bibfnamefont {Z.}~\bibnamefont
			{Chen}}, \bibinfo {author} {\bibfnamefont {J.}~\bibnamefont {Kelly}},
		\bibinfo {author} {\bibfnamefont {C.}~\bibnamefont {Quintana}}, \bibinfo
		{author} {\bibfnamefont {R.}~\bibnamefont {Barends}}, \bibinfo {author}
		{\bibfnamefont {B.}~\bibnamefont {Campbell}}, \bibinfo {author}
		{\bibfnamefont {Yu}~\bibnamefont {Chen}}, \bibinfo {author} {\bibfnamefont
			{B.}~\bibnamefont {Chiaro}}, \bibinfo {author} {\bibfnamefont
			{A.}~\bibnamefont {Dunsworth}}, \bibinfo {author} {\bibfnamefont {A.~G.}\
			\bibnamefont {Fowler}}, \bibinfo {author} {\bibfnamefont {E.}~\bibnamefont
			{Jeffrey}}, \bibinfo {author} {\bibfnamefont {A.}~\bibnamefont {Megrant}},
		\bibinfo {author} {\bibfnamefont {J.}~\bibnamefont {Mutus}}, \bibinfo
		{author} {\bibfnamefont {C.}~\bibnamefont {Neill}}, \bibinfo {author}
		{\bibfnamefont {P.~J.~J.}\ \bibnamefont {O'Malley}}, \bibinfo {author}
		{\bibfnamefont {P.}~\bibnamefont {Roushan}}, \bibinfo {author} {\bibfnamefont
			{D.}~\bibnamefont {Sank}}, \bibinfo {author} {\bibfnamefont {A.}~\bibnamefont
			{Vainsencher}}, \bibinfo {author} {\bibfnamefont {J.}~\bibnamefont {Wenner}},
		\bibinfo {author} {\bibfnamefont {T.~C.}\ \bibnamefont {White}}, \bibinfo
		{author} {\bibfnamefont {A.~N.}\ \bibnamefont {Cleland}}, \ and\ \bibinfo
		{author} {\bibfnamefont {J.~M.}\ \bibnamefont {Martinis}},\ }\bibfield
	{title} {\enquote {\bibinfo {title} {Measuring and suppressing quantum state
				leakage in a superconducting qubit},}\ }\href@noop {} {\bibfield  {journal}
		{\bibinfo  {journal} {Phys. Rev. Lett.}\ }\textbf {\bibinfo {volume} {116}},\
		\bibinfo {pages} {020501} (\bibinfo {year} {2016})}\BibitemShut {NoStop}%
	\bibitem [{\citenamefont {Rol}\ \emph {et~al.}(2019)\citenamefont {Rol},
		\citenamefont {Battistel}, \citenamefont {Malinowski}, \citenamefont
		{Bultink}, \citenamefont {Tarasinski}, \citenamefont {Vollmer}, \citenamefont
		{Haider}, \citenamefont {Muthusubramanian}, \citenamefont {Bruno},
		\citenamefont {Terhal},\ and\ \citenamefont {DiCarlo}}]{rol2019fast}%
	\BibitemOpen
	\bibfield  {author} {\bibinfo {author} {\bibfnamefont {M.~A.}\ \bibnamefont
			{Rol}}, \bibinfo {author} {\bibfnamefont {F.}~\bibnamefont {Battistel}},
		\bibinfo {author} {\bibfnamefont {F.~K.}\ \bibnamefont {Malinowski}},
		\bibinfo {author} {\bibfnamefont {C.~C.}\ \bibnamefont {Bultink}}, \bibinfo
		{author} {\bibfnamefont {B.~M.}\ \bibnamefont {Tarasinski}}, \bibinfo
		{author} {\bibfnamefont {R.}~\bibnamefont {Vollmer}}, \bibinfo {author}
		{\bibfnamefont {N.}~\bibnamefont {Haider}}, \bibinfo {author} {\bibfnamefont
			{N.}~\bibnamefont {Muthusubramanian}}, \bibinfo {author} {\bibfnamefont
			{A.}~\bibnamefont {Bruno}}, \bibinfo {author} {\bibfnamefont {B.~M.}\
			\bibnamefont {Terhal}}, \ and\ \bibinfo {author} {\bibfnamefont
			{L.}~\bibnamefont {DiCarlo}},\ }\bibfield  {title} {\enquote {\bibinfo
			{title} {A fast, low-leakage, high-fidelity two-qubit gate for a programmable
				superconducting quantum computer},}\ }\href@noop {} {\bibfield  {journal}
		{\bibinfo  {journal} {Phys. Rev. Lett.}\ }\textbf {\bibinfo {volume} {123}},\
		\bibinfo {pages} {120502} (\bibinfo {year} {2019})}\BibitemShut {NoStop}%
	\bibitem [{\citenamefont {Clerk}\ \emph {et~al.}(2010)\citenamefont {Clerk},
		\citenamefont {Devoret}, \citenamefont {Girvin}, \citenamefont {Marquardt},\
		and\ \citenamefont {Schoelkopf}}]{clerk2010introduction}%
	\BibitemOpen
	\bibfield  {author} {\bibinfo {author} {\bibfnamefont {Aashish~A}\
			\bibnamefont {Clerk}}, \bibinfo {author} {\bibfnamefont {Michel~H}\
			\bibnamefont {Devoret}}, \bibinfo {author} {\bibfnamefont {Steven~M}\
			\bibnamefont {Girvin}}, \bibinfo {author} {\bibfnamefont {Florian}\
			\bibnamefont {Marquardt}}, \ and\ \bibinfo {author} {\bibfnamefont
			{Robert~J}\ \bibnamefont {Schoelkopf}},\ }\bibfield  {title} {\enquote
		{\bibinfo {title} {Introduction to quantum noise, measurement, and
				amplification},}\ }\href@noop {} {\bibfield  {journal} {\bibinfo  {journal}
			{Reviews of Modern Physics}\ }\textbf {\bibinfo {volume} {82}},\ \bibinfo
		{pages} {1155--1208} (\bibinfo {year} {2010})}\BibitemShut {NoStop}%
	\bibitem [{\citenamefont {Wang}\ \emph {et~al.}(2019)\citenamefont {Wang},
		\citenamefont {Shankar}, \citenamefont {Minev}, \citenamefont
		{Campagne-Ibarcq}, \citenamefont {Narla},\ and\ \citenamefont
		{Devoret}}]{wang2019cavity}%
	\BibitemOpen
	\bibfield  {author} {\bibinfo {author} {\bibfnamefont {Z}~\bibnamefont
			{Wang}}, \bibinfo {author} {\bibfnamefont {S}~\bibnamefont {Shankar}},
		\bibinfo {author} {\bibfnamefont {ZK}~\bibnamefont {Minev}}, \bibinfo
		{author} {\bibfnamefont {Philippe}\ \bibnamefont {Campagne-Ibarcq}}, \bibinfo
		{author} {\bibfnamefont {A}~\bibnamefont {Narla}}, \ and\ \bibinfo {author}
		{\bibfnamefont {Michel~H}\ \bibnamefont {Devoret}},\ }\bibfield  {title}
	{\enquote {\bibinfo {title} {Cavity attenuators for superconducting
				qubits},}\ }\href@noop {} {\bibfield  {journal} {\bibinfo  {journal}
			{Physical Review Applied}\ }\textbf {\bibinfo {volume} {11}},\ \bibinfo
		{pages} {014031} (\bibinfo {year} {2019})}\BibitemShut {NoStop}%
	\bibitem [{\citenamefont {Frisk~Kockum}\ \emph {et~al.}(2019)\citenamefont
		{Frisk~Kockum}, \citenamefont {Miranowicz}, \citenamefont {De~Liberato},
		\citenamefont {Savasta},\ and\ \citenamefont {Nori}}]{frisk2019ultrastrong}%
	\BibitemOpen
	\bibfield  {author} {\bibinfo {author} {\bibfnamefont {Anton}\ \bibnamefont
			{Frisk~Kockum}}, \bibinfo {author} {\bibfnamefont {Adam}\ \bibnamefont
			{Miranowicz}}, \bibinfo {author} {\bibfnamefont {Simone}\ \bibnamefont
			{De~Liberato}}, \bibinfo {author} {\bibfnamefont {Salvatore}\ \bibnamefont
			{Savasta}}, \ and\ \bibinfo {author} {\bibfnamefont {Franco}\ \bibnamefont
			{Nori}},\ }\bibfield  {title} {\enquote {\bibinfo {title} {Ultrastrong
				coupling between light and matter},}\ }\href@noop {} {\bibfield  {journal}
		{\bibinfo  {journal} {Nature Reviews Physics}\ }\textbf {\bibinfo {volume}
			{1}},\ \bibinfo {pages} {19--40} (\bibinfo {year} {2019})}\BibitemShut
	{NoStop}%
	\bibitem [{\citenamefont {Egger}\ and\ \citenamefont
		{Wilhelm}(2013)}]{egger2013multimode}%
	\BibitemOpen
	\bibfield  {author} {\bibinfo {author} {\bibfnamefont {Daniel~J}\
			\bibnamefont {Egger}}\ and\ \bibinfo {author} {\bibfnamefont {Frank~K}\
			\bibnamefont {Wilhelm}},\ }\bibfield  {title} {\enquote {\bibinfo {title}
			{Multimode circuit quantum electrodynamics with hybrid metamaterial
				transmission lines},}\ }\href@noop {} {\bibfield  {journal} {\bibinfo
			{journal} {Physical review letters}\ }\textbf {\bibinfo {volume} {111}},\
		\bibinfo {pages} {163601} (\bibinfo {year} {2013})}\BibitemShut {NoStop}%
	\bibitem [{\citenamefont {Khaneja}\ \emph {et~al.}(2005)\citenamefont
		{Khaneja}, \citenamefont {Reiss}, \citenamefont {Kehlet}, \citenamefont
		{Schulte-Herbr{\"u}ggen},\ and\ \citenamefont {Glaser}}]{khaneja2005optimal}%
	\BibitemOpen
	\bibfield  {author} {\bibinfo {author} {\bibfnamefont {Navin}\ \bibnamefont
			{Khaneja}}, \bibinfo {author} {\bibfnamefont {Timo}\ \bibnamefont {Reiss}},
		\bibinfo {author} {\bibfnamefont {Cindie}\ \bibnamefont {Kehlet}}, \bibinfo
		{author} {\bibfnamefont {Thomas}\ \bibnamefont {Schulte-Herbr{\"u}ggen}}, \
		and\ \bibinfo {author} {\bibfnamefont {Steffen~J}\ \bibnamefont {Glaser}},\
	}\bibfield  {title} {\enquote {\bibinfo {title} {Optimal control of coupled
				spin dynamics: design of nmr pulse sequences by gradient ascent
				algorithms},}\ }\href@noop {} {\bibfield  {journal} {\bibinfo  {journal}
			{Journal of magnetic resonance}\ }\textbf {\bibinfo {volume} {172}},\
		\bibinfo {pages} {296--305} (\bibinfo {year} {2005})}\BibitemShut {NoStop}%
	\bibitem [{\citenamefont {Machnes}\ \emph {et~al.}(2011)\citenamefont
		{Machnes}, \citenamefont {Sander}, \citenamefont {Glaser}, \citenamefont
		{de~Fouquieres}, \citenamefont {Gruslys}, \citenamefont {Schirmer},\ and\
		\citenamefont {Schulte-Herbr{\"u}ggen}}]{machnes2011comparing}%
	\BibitemOpen
	\bibfield  {author} {\bibinfo {author} {\bibfnamefont {Shai}\ \bibnamefont
			{Machnes}}, \bibinfo {author} {\bibfnamefont {Urgen}\ \bibnamefont {Sander}},
		\bibinfo {author} {\bibfnamefont {Steffen~J}\ \bibnamefont {Glaser}},
		\bibinfo {author} {\bibfnamefont {Pierre}\ \bibnamefont {de~Fouquieres}},
		\bibinfo {author} {\bibfnamefont {Audrunas}\ \bibnamefont {Gruslys}},
		\bibinfo {author} {\bibfnamefont {Sonia}\ \bibnamefont {Schirmer}}, \ and\
		\bibinfo {author} {\bibfnamefont {Thomas}\ \bibnamefont
			{Schulte-Herbr{\"u}ggen}},\ }\bibfield  {title} {\enquote {\bibinfo {title}
			{Comparing, optimizing, and benchmarking quantum-control algorithms in a
				unifying programming framework},}\ }\href@noop {} {\bibfield  {journal}
		{\bibinfo  {journal} {Physical Review A—Atomic, Molecular, and Optical
				Physics}\ }\textbf {\bibinfo {volume} {84}},\ \bibinfo {pages} {022305}
		(\bibinfo {year} {2011})}\BibitemShut {NoStop}%
	\bibitem [{\citenamefont {Sivak}\ \emph {et~al.}(2022)\citenamefont {Sivak},
		\citenamefont {Eickbusch}, \citenamefont {Liu}, \citenamefont {Royer},
		\citenamefont {Tsioutsios},\ and\ \citenamefont {Devoret}}]{sivak2022model}%
	\BibitemOpen
	\bibfield  {author} {\bibinfo {author} {\bibfnamefont {VV}~\bibnamefont
			{Sivak}}, \bibinfo {author} {\bibfnamefont {A}~\bibnamefont {Eickbusch}},
		\bibinfo {author} {\bibfnamefont {H}~\bibnamefont {Liu}}, \bibinfo {author}
		{\bibfnamefont {B}~\bibnamefont {Royer}}, \bibinfo {author} {\bibfnamefont
			{I}~\bibnamefont {Tsioutsios}}, \ and\ \bibinfo {author} {\bibfnamefont
			{MH}~\bibnamefont {Devoret}},\ }\bibfield  {title} {\enquote {\bibinfo
			{title} {Model-free quantum control with reinforcement learning},}\
	}\href@noop {} {\bibfield  {journal} {\bibinfo  {journal} {Physical Review
				X}\ }\textbf {\bibinfo {volume} {12}},\ \bibinfo {pages} {011059} (\bibinfo
		{year} {2022})}\BibitemShut {NoStop}%
	\bibitem [{\citenamefont {F{\"o}sel}\ \emph {et~al.}(2018)\citenamefont
		{F{\"o}sel}, \citenamefont {Tighineanu}, \citenamefont {Weiss},\ and\
		\citenamefont {Marquardt}}]{fosel2018reinforcement}%
	\BibitemOpen
	\bibfield  {author} {\bibinfo {author} {\bibfnamefont {Thomas}\ \bibnamefont
			{F{\"o}sel}}, \bibinfo {author} {\bibfnamefont {Petru}\ \bibnamefont
			{Tighineanu}}, \bibinfo {author} {\bibfnamefont {Talitha}\ \bibnamefont
			{Weiss}}, \ and\ \bibinfo {author} {\bibfnamefont {Florian}\ \bibnamefont
			{Marquardt}},\ }\bibfield  {title} {\enquote {\bibinfo {title} {Reinforcement
				learning with neural networks for quantum feedback},}\ }\href@noop {}
	{\bibfield  {journal} {\bibinfo  {journal} {Physical Review X}\ }\textbf
		{\bibinfo {volume} {8}},\ \bibinfo {pages} {031084} (\bibinfo {year}
		{2018})}\BibitemShut {NoStop}%
	\bibitem [{\citenamefont {Bukov}\ \emph {et~al.}(2018)\citenamefont {Bukov},
		\citenamefont {Day}, \citenamefont {Sels}, \citenamefont {Weinberg},
		\citenamefont {Polkovnikov},\ and\ \citenamefont
		{Mehta}}]{bukov2018reinforcement}%
	\BibitemOpen
	\bibfield  {author} {\bibinfo {author} {\bibfnamefont {Marin}\ \bibnamefont
			{Bukov}}, \bibinfo {author} {\bibfnamefont {Alexandre~GR}\ \bibnamefont
			{Day}}, \bibinfo {author} {\bibfnamefont {Dries}\ \bibnamefont {Sels}},
		\bibinfo {author} {\bibfnamefont {Phillip}\ \bibnamefont {Weinberg}},
		\bibinfo {author} {\bibfnamefont {Anatoli}\ \bibnamefont {Polkovnikov}}, \
		and\ \bibinfo {author} {\bibfnamefont {Pankaj}\ \bibnamefont {Mehta}},\
	}\bibfield  {title} {\enquote {\bibinfo {title} {Reinforcement learning in
				different phases of quantum control},}\ }\href@noop {} {\bibfield  {journal}
		{\bibinfo  {journal} {Physical Review X}\ }\textbf {\bibinfo {volume} {8}},\
		\bibinfo {pages} {031086} (\bibinfo {year} {2018})}\BibitemShut {NoStop}%
	\bibitem [{\citenamefont {Niu}\ \emph {et~al.}(2023)\citenamefont {Niu},
		\citenamefont {Zhang}, \citenamefont {Liu}, \citenamefont {Qiu},
		\citenamefont {Huang}, \citenamefont {Huang}, \citenamefont {Jia},
		\citenamefont {Liu}, \citenamefont {Tao}, \citenamefont {Wei} \emph
		{et~al.}}]{niu2023low}%
	\BibitemOpen
	\bibfield  {author} {\bibinfo {author} {\bibfnamefont {Jingjing}\
			\bibnamefont {Niu}}, \bibinfo {author} {\bibfnamefont {Libo}\ \bibnamefont
			{Zhang}}, \bibinfo {author} {\bibfnamefont {Yang}\ \bibnamefont {Liu}},
		\bibinfo {author} {\bibfnamefont {Jiawei}\ \bibnamefont {Qiu}}, \bibinfo
		{author} {\bibfnamefont {Wenhui}\ \bibnamefont {Huang}}, \bibinfo {author}
		{\bibfnamefont {Jiaxiang}\ \bibnamefont {Huang}}, \bibinfo {author}
		{\bibfnamefont {Hao}\ \bibnamefont {Jia}}, \bibinfo {author} {\bibfnamefont
			{Jiawei}\ \bibnamefont {Liu}}, \bibinfo {author} {\bibfnamefont {Ziyu}\
			\bibnamefont {Tao}}, \bibinfo {author} {\bibfnamefont {Weiwei}\ \bibnamefont
			{Wei}},  \emph {et~al.},\ }\bibfield  {title} {\enquote {\bibinfo {title}
			{Low-loss interconnects for modular superconducting quantum processors},}\
	}\href@noop {} {\bibfield  {journal} {\bibinfo  {journal} {Nature
				Electronics}\ }\textbf {\bibinfo {volume} {6}},\ \bibinfo {pages} {235--241}
		(\bibinfo {year} {2023})}\BibitemShut {NoStop}%
\end{thebibliography}
\end{document}